%%%%%%%%%%%%%%%%%%%%%%% PLAIN TEX FILE %%%%%%%%%%%%%%%%%%%

\newcount\mgnf\newcount\tipi\newcount\tipoformule\newcount\greco

\tipi=2          %uso caratteri: 2=cmcompleti, 1=cmparziali, 0=amparziali
\tipoformule=0   %=0 da numeroparagrafo.numeroformula; se no numero
                 %assoluto

\global\newcount\numsec
\global\newcount\numfor
\global\newcount\numtheo
\global\advance\numtheo by 1

\def\senondefinito#1{\expandafter\ifx\csname#1\endcsname\relax}

\def\SIA #1,#2,#3 {\senondefinito{#1#2}%
\expandafter\xdef\csname #1#2\endcsname{#3}\else
\write16{???? ma #1,#2 e' gia' stato definito !!!!} \fi}

\def\etichetta(#1){(\veroparagrafo.\veraformula)%
\SIA e,#1,(\veroparagrafo.\veraformula) %
\global\advance\numfor by 1%
\write15{\string\FU (#1){\equ(#1)}}%
\write16{ EQ #1 ==> \equ(#1) }}

\def\letichetta(#1){\veroparagrafo.\verotheo
\SIA e,#1,{\veroparagrafo.\verotheo}
\global\advance\numtheo by 1
\write15{\string\FU (#1){\equ(#1)}}
\write16{ Sta \equ(#1) == #1 }}

\def\tetichetta(#1){\veroparagrafo.\veraformula %%%%copy four lines
\SIA e,#1,{(\veroparagrafo.\veraformula)}
\global\advance\numfor by 1
\write15{\string\FU (#1){\equ(#1)}}
\write16{ tag #1 ==> \equ(#1)}}

\def\FU(#1)#2{\SIA fu,#1,#2 }

\def\etichettaa(#1){(A\veroparagrafo.\veraformula)%
\SIA e,#1,(A\veroparagrafo.\veraformula) %
\global\advance\numfor by 1%
\write15{\string\FU (#1){\equ(#1)}}%
\write16{ EQ #1 ==> \equ(#1) }}

\def\BOZZA{
\def\alato(##1){%
 {\rlap{\kern-\hsize\kern-1.4truecm{$\scriptstyle##1$}}}}%
\def\aolado(##1){%
 {%\vtop to \profonditastruttura
{%\baselineskip
 %\profonditastruttura\vss
 \rlap{\kern-1.4truecm{$\scriptstyle##1$}}}}}
}

\def\alato(#1){}
\def\aolado(#1){}

\def\veroparagrafo{\number\numsec}
\def\veraformula{\number\numfor}
\def\verotheo{\number\numtheo}

\def\Eq(#1){\eqno{\etichetta(#1)\alato(#1)}}
\def\eq(#1){\etichetta(#1)\alato(#1)}
\def\leq(#1){\leqno{\aolado(#1)\etichetta(#1)}}%%%%%this line for \leqno
\def\teq(#1){\tag{\aolado(#1)\tetichetta(#1)\alato(#1)}}%%%%%this line for\tag
\def\Eqa(#1){\eqno{\etichettaa(#1)\alato(#1)}}
\def\eqa(#1){\etichettaa(#1)\alato(#1)}
\def\eqv(#1){\senondefinito{fu#1}$\clubsuit$#1
\write16{#1 non e' (ancora) definito}%
\else\csname fu#1\endcsname\fi}
\def\equ(#1){\senondefinito{e#1}\eqv(#1)\else\csname e#1\endcsname\fi}

%%%% next six lines by paf (no responsibilities taken)
\def\Lemma(#1){\aolado(#1)Lemma \letichetta(#1)}%
\def\Theorem(#1){{\aolado(#1)Theorem \letichetta(#1)}}%
\def\Proposition(#1){\aolado(#1){Proposition \letichetta(#1)}}%
\def\Corollary(#1){{\aolado(#1)Corollary \letichetta(#1)}}%
\def\Remark(#1){{\noindent\aolado(#1){\bf Remark \letichetta(#1).}}}%
\def\Definition(#1){{\noindent\aolado(#1){\bf Definition 
\letichetta(#1)$\!\!$\hskip-1.6truemm}}}
\def\Example(#1){\aolado(#1) Example \letichetta(#1)$\!\!$\hskip-1.6truemm}

\def\include#1{
\openin13=#1.aux \ifeof13 \relax \else
\input #1.aux \closein13 \fi}

\openin14=\jobname.aux \ifeof14 \relax \else
\input \jobname.aux \closein14 \fi
\openout15=\jobname.aux

%%%%%%%%%%%%%%%%%%%%%%%%%%%%%%%%%%%%%%%%%%%%%%
%%%%%%%%%%%%%%%%%%%%%  Numerazione pagine

{\count255=\time\divide\count255 by 60 \xdef\hourmin{\number\count255}
        \multiply\count255 by-60\advance\count255 by\time
   \xdef\hourmin{\hourmin:\ifnum\count255<10 0\fi\the\count255}}

\def\oramin{\hourmin }

\def\data{\number\day/\ifcase\month\or january \or february \or march \or
april \or may \or june \or july \or august \or september
\or october \or november \or december \fi/\number\year;\ \oramin}

\newcount\pgn \pgn=1
\def\foglio{\number\numsec:\number\pgn
\global\advance\pgn by 1}
\def\foglioa{A\number\numsec:\number\pgn
\global\advance\pgn by 1}

\footline={\rlap{\hbox{\copy200}}\hss\tenrm\folio\hss}
%\footline={\hss\tenrm\folio\hss}

%%%%%%%%
%% am
\def\TIPIO{
\font\setterm=amr7 %\font\settei=ammi7
%\font\settesy=amsy7 \font\settebf=ambx7 %\font\setteit=amit7
%%%%% cambiamenti di formato %%%
\def \settepunti{\def\rm{\fam0\setterm}% passaggio a tipi da 7-punti
\textfont0=\setterm   %\textfont1=\settei
%\textfont2=\settesy   %\textfont3=\setteit
%\textfont\itfam=\setteit  \def\it{\fam\itfam\setteit}
%\textfont\bffam=\settebf  \def\bf{\fam\bffam\settebf}
\normalbaselineskip=9pt\normalbaselines\rm
}\let\nota=\settepunti}
%%%%%%%

%%cm completo
\def\TIPITOT{
\font\twelverm=cmr12
\font\twelvei=cmmi12
\font\twelvesy=cmsy10 scaled\magstep1
\font\twelveex=cmex10 scaled\magstep1
\font\twelveit=cmti12
\font\twelvett=cmtt12
\font\twelvebf=cmbx12 scaled\magstep1
\font\twelvesl=cmsl12
\font\ninerm=cmr9
\font\ninesy=cmsy9
\font\eightrm=cmr8
\font\eighti=cmmi8
\font\eightsy=cmsy8
\font\eightbf=cmbx8
\font\eighttt=cmtt8
\font\eightsl=cmsl8
\font\eightit=cmti8
\font\sixrm=cmr6
\font\sixbf=cmbx6
\font\sixi=cmmi6
\font\sixsy=cmsy6
%%%%%%%%%%%%%%%%%%%%%%%%%%%%%%%%%%%%%%%
\font\twelvetruecmr=cmr10 scaled\magstep1
\font\twelvetruecmsy=cmsy10 scaled\magstep1
\font\tentruecmr=cmr10
\font\tentruecmsy=cmsy10
\font\eighttruecmr=cmr8
\font\eighttruecmsy=cmsy8
\font\seventruecmr=cmr7
\font\seventruecmsy=cmsy7
\font\sixtruecmr=cmr6
\font\sixtruecmsy=cmsy6
\font\fivetruecmr=cmr5
\font\fivetruecmsy=cmsy5
%%%% definizioni per 10pt %%%%%%%%
\textfont\truecmr=\tentruecmr
\scriptfont\truecmr=\seventruecmr
\scriptscriptfont\truecmr=\fivetruecmr
\textfont\truecmsy=\tentruecmsy
\scriptfont\truecmsy=\seventruecmsy
\scriptscriptfont\truecmr=\fivetruecmr
\scriptscriptfont\truecmsy=\fivetruecmsy
%%%%% cambio grandezza %%%%%%
\def \eightpoint{\def\rm{\fam0\eightrm}% switch to 8-point type
\textfont0=\eightrm \scriptfont0=\sixrm \scriptscriptfont0=\fiverm
\textfont1=\eighti \scriptfont1=\sixi   \scriptscriptfont1=\fivei
\textfont2=\eightsy \scriptfont2=\sixsy   \scriptscriptfont2=\fivesy
\textfont3=\tenex \scriptfont3=\tenex   \scriptscriptfont3=\tenex
\textfont\itfam=\eightit  \def\it{\fam\itfam\eightit}%
\textfont\slfam=\eightsl  \def\sl{\fam\slfam\eightsl}%
\textfont\ttfam=\eighttt  \def\tt{\fam\ttfam\eighttt}%
\textfont\bffam=\eightbf  \scriptfont\bffam=\sixbf
\scriptscriptfont\bffam=\fivebf  \def\bf{\fam\bffam\eightbf}%
\tt \ttglue=.5em plus.25em minus.15em
\setbox\strutbox=\hbox{\vrule height7pt depth2pt width0pt}%
\normalbaselineskip=9pt
\let\sc=\sixrm  \let\big=\eightbig  \normalbaselines\rm
\textfont\truecmr=\eighttruecmr
\scriptfont\truecmr=\sixtruecmr
\scriptscriptfont\truecmr=\fivetruecmr
\textfont\truecmsy=\eighttruecmsy
\scriptfont\truecmsy=\sixtruecmsy
}\let\nota=\eightpoint}

\newfam\msbfam   %per uso in \TIPITOT
\newfam\truecmr  %per uso in \TIPITOT
\newfam\truecmsy %per uso in \TIPITOT
%%%%%%%%%%%%%%%%%%%%%%%%%%%%%%%
%%Scelta dei caratteri
%\newcount\tipi \tipi=0   %e' definito all'inizio
\newskip\ttglue
\ifnum\tipi=0\TIPIO \else\ifnum\tipi=1 \TIPI\else \TIPITOT\fi\fi

\def\a{\alpha}
\def\b{\beta}
\def\d{\delta}
\def\e{\epsilon}
\def\ve{\varepsilon}

\def\vf{\varphi}
\def\g{\gamma}

\def\l{\lambda}

\def\s{\sigma}
\def\t{\tau}

\def\o{\omega}

\def\L{\Lambda}

\def\O{\Omega}

\def\E{{I\kern-.25em{E}}}
\def\N{{I\kern-.25em{N}}}
\def\M{{I\kern-.25em{M}}}
\def\R{{I\kern-.25em{R}}}
\def\Z{{Z\kern-.425em{Z}}}
\def\1{{1\kern-.25em\hbox{\rm I}}}
\def\eu{{1\kern-.25em\hbox{\sm I}}}

\def\C{{I\kern-.64em{C}}}
\def\P{{I\kern-.25em{P}}}

%\def\P{\hskip.2em\hbox{\rm P\kern-0.8em{I}\hskip.7em}}

% Spezielle Definitionen

\def\FF{{\cal F}}
\def\GG{{\cal G}}
\def\HH{{\cal H}}

\def\TT{{\cal T}}

\def\MM{{\cal M}}
\def\WW{{\cal W}}

\def\UU{{\cal U}}
\def\LL{{\cal L}}
\def\XX{{\cal X}}

\def\sign{\,\hbox{sign}\,}

\def\sqr#1#2{{\vcenter{\vbox{\hrule height.#2pt
     \hbox{\vrule width.#2pt height#1pt \kern#1pt
   \vrule width.#2pt}\hrule height.#2pt}}}}
\def\qed{ $\mathchoice\sqr64\sqr64\sqr{2.1}3\sqr{1.5}3$} 

%   Non-character macros

\newcount\foot
\foot=1
\def\note#1{\footnote{${}^{\number\foot}$}{\ftn #1}\advance\foot by 1}
\def\tag #1{\eqno{\hbox{\rm(#1)}}}
\def\frac#1#2{{#1\over #2}}

\def\text#1{\quad{\hbox{#1}}\quad}
\def\newpage{\vfill\eject}

\def\thanks{\noindent{\bf Aknowledgements: }}

%\font\thbf=cmcsc10 scaled\magstep1

% Font-Definitions

\font\ftn=cmr8

\font\it=cmti10
\font\bf=cmbx10
\font\sm=cmr7

%%%%%%%%References macros start here%%%%%%%%%%
%
\catcode`\X=12\catcode`\@=11
\def\n@wcount{\alloc@0\count\countdef\insc@unt}
\def\n@wwrite{\alloc@7\write\chardef\sixt@@n}
\def\n@wread{\alloc@6\read\chardef\sixt@@n}
\def\crossrefs#1{\ifx\alltgs#1\let\tr@ce=\alltgs\else\def\tr@ce{#1,}\fi
   \n@wwrite\cit@tionsout\openout\cit@tionsout=\jobname.cit 
   \write\cit@tionsout{\tr@ce}\expandafter\setfl@gs\tr@ce,}
\def\setfl@gs#1,{\def\@{#1}\ifx\@\empty\let\next=\relax
   \else\let\next=\setfl@gs\expandafter\xdef
   \csname#1tr@cetrue\endcsname{}\fi\next}
\newcount\sectno\sectno=0\newcount\subsectno\subsectno=0\def\r@s@t{\relax}
\def\resetall{\global\advance\sectno by 1\subsectno=0
  \gdef\firstpart{\number\sectno}\r@s@t}
\def\resetsub{\global\advance\subsectno by 1
   \gdef\firstpart{\number\sectno.\number\subsectno}\r@s@t}
\def\v@idline{\par}\def\firstpart{\number\sectno}
\def\l@c@l#1X{\firstpart.#1}\def\gl@b@l#1X{#1}\def\t@d@l#1X{{}}
\def\m@ketag#1#2{\expandafter\n@wcount\csname#2tagno\endcsname
     \csname#2tagno\endcsname=0\let\tail=\alltgs\xdef\alltgs{\tail#2,}%
  \ifx#1\l@c@l\let\tail=\r@s@t\xdef\r@s@t{\csname#2tagno\endcsname=0\tail}\fi
   \expandafter\gdef\csname#2cite\endcsname##1{\expandafter
 %the following line was replaced by the subseqent one, DNA 7/6/89
  %  \ifx\csname#2tag##1\endcsname\relax?\else\csname#2tag##1\endcsname\fi
     \ifx\csname#2tag##1\endcsname\relax?\else{\rm\csname#2tag##1\endcsname}\fi
    \expandafter\ifx\csname#2tr@cetrue\endcsname\relax\else
     \write\cit@tionsout{#2tag ##1 cited on page \folio.}\fi}%
   \expandafter\gdef\csname#2page\endcsname##1{\expandafter
     \ifx\csname#2page##1\endcsname\relax?\else\csname#2page##1\endcsname\fi
     \expandafter\ifx\csname#2tr@cetrue\endcsname\relax\else
     \write\cit@tionsout{#2tag ##1 cited on page \folio.}\fi}%
   \expandafter\gdef\csname#2tag\endcsname##1{\global\advance
     \csname#2tagno\endcsname by 1%
   \expandafter\ifx\csname#2check##1\endcsname\relax\else%
\fi%      \immediate\write16{Warning: #2tag ##1 used more than once.}\fi
   \expandafter\xdef\csname#2check##1\endcsname{}%
   \expandafter\xdef\csname#2tag##1\endcsname
     {#1\number\csname#2tagno\endcsnameX}%
   \write\t@gsout{#2tag ##1 assigned number \csname#2tag##1\endcsname\space
      on page \number\count0.}%
   \csname#2tag##1\endcsname}}%
\def\m@kecs #1tag #2 assigned number #3 on page #4.%
   {\expandafter\gdef\csname#1tag#2\endcsname{#3}
   \expandafter\gdef\csname#1page#2\endcsname{#4}}
\def\re@der{\ifeof\t@gsin\let\next=\relax\else
    \read\t@gsin to\t@gline\ifx\t@gline\v@idline\else
    \expandafter\m@kecs \t@gline\fi\let \next=\re@der\fi\next}
\def\t@gs#1{\def\alltgs{}\m@ketag#1e\m@ketag#1s\m@ketag\t@d@l p
    \m@ketag\gl@b@l r \n@wread\t@gsin\openin\t@gsin=\jobname.tgs \re@der
    \closein\t@gsin\n@wwrite\t@gsout\openout\t@gsout=\jobname.tgs }
\outer\def\localtags{\t@gs\l@c@l}
\outer\def\globaltags{\t@gs\gl@b@l}
\outer\def\newlocaltag#1{\m@ketag\l@c@l{#1}}
\outer\def\newglobaltag#1{\m@ketag\gl@b@l{#1}}

\def\t@gsoff#1,{\def\@{#1}\ifx\@\empty\let\next=\relax\else\let\next=\t@gsoff
   \expandafter\gdef\csname#1cite\endcsname{\relax}
   \expandafter\gdef\csname#1page\endcsname##1{?}
   \expandafter\gdef\csname#1tag\endcsname{\relax}\fi\next}
\def\verbatimtags{\let\ift@gs=\iffalse\ifx\alltgs\relax\else
   \expandafter\t@gsoff\alltgs,\fi}
\catcode`\X=11 \catcode`\@=\active
\localtags
%%%%%%%%%%%%references macro end here%%%%%%%%%%%%%%%
%
%%%%%%%%%%%%%%%%%end of macros%%%%%%%%%%%%%%%%%%%%%%

%%\setbox200\hbox{$\scriptscriptstyle \data $}

\global\newcount\numpunt
\magnification=\magstep1

%\voffset=.8truecm
%\hoffset=-.2truecm

\baselineskip=14pt  
\parindent=12pt
\lineskip=4pt\lineskiplimit=0.1pt
\parskip=0.1pt plus1pt
\hfuzz=1pt

\hyphenation{small}

\centerline {\twelvebf
Hydrodynamic Limit of Brownian Particles Interacting}
\vskip.3truecm
\centerline{\twelvebf with Short and Long Range Forces}

\vskip1cm
\centerline{ 
Paolo Butt\`a \footnote{$^1$}{\eightrm Department of Mathematics, 
Rutgers, the State University of New Jersey,}
\footnote{}{\eightrm 110 Frelinghuysen rd, Piscataway NJ 08854--8019, 
USA.}$^3$,\hskip.2cm
Joel L. Lebowitz \footnote{$^2$} {\eightrm Departments of Mathematics
and Physics, Rutgers, the State University of New Jersey,}
\footnote{}{\eightrm 110 Frelinghuysen rd, Piscataway NJ 08854--8019, 
USA.}\footnote{$^3$}{\eightrm The work was supported in part by 
NSF Grant 95--23266 and AFOSR 95--0159. P. B. is visiting}}
\footnote{}{\eightrm Rutgers University with a CNR  
(Italian National Research Council) Fellowship in Mathematics.}
\footnote{}{\eightrm {\eightit Key Words}: Interacting Particle
Systems, Hydrodynamic Limit, Non-local Evolution Equations.}
\footnote{}{\eightrm {\eightit 1991 Mathematics Subject Classification}: 
45K05, 60F10, 60J60, 60K35, 82C22.} 

\vskip.5cm 
\centerline{Rutgers University, New Jersey, USA}
\vskip1truecm

\centerline{\bf Abstract.}
\vskip.5truecm

We investigate the time evolution of a model system of interacting
particles, moving in a $d$-dimensional torus. The microscopic dynamics
are first order in time with velocities set equal to the negative
gradient of a potential energy term $\Psi$ plus independent Brownian
motions: $\Psi$ is the sum of pair potentials, $V(r)+\g^{d}J(\g r)$,
the second term has the form of a Kac potential with inverse range
$\g$. Using diffusive hydrodynamical scaling (spatial scale $\g^{-1}$,
temporal scale $\g^{-2}$) we obtain, in the limit $\g\downarrow 0$, 
a diffusive type integro-differential equation describing the time
evolution of the macroscopic density profile.

\newpage
\goodbreak 
\vskip1truecm
\centerline {\bf 1. Introduction.}
\vskip.5truecm
\numsec= 1
\numfor= 1
\numtheo=1

The transition from the microscopic dynamics of interacting particles
to hydrodynamical type equations describing the coarse grained
evolution of macroscopic variables, such as the diffusion equation 
for the density, is a basic problem of non-equilibrium statistical 
mechanics. While far from resolved for systems with realistic 
interactions there has been much progress recently on this problem for
model systems. Like in real systems, the transition from microscopic
to macroscopic evolutions in these models is based on a separation
between microscopic and macroscopic scales. Setting $\ve$ equal to the
ratio of microscopic to macroscopic spatial scale and then looking at
macroscopic times which are of order $\ve^{-\a}$ microscopic time
units, $\a=1$ for Euler (non-dissipative) and $\a=2$ for diffusive 
evolutions, we expect to obtain the macroscopic equations in the 
hydrodynamical scaling limit (HSL) $\ve\downarrow 0$. We refer to
the books of De Masi and Presutti, [\rcite{DP}], and Spohn,
[\rcite{S}], for a general background on this subject (see also the
review article by Lebowitz, Presutti, and Spohn, [\rcite{LPS}]).

To actually prove this HSL, one needs to show that during macroscopic
evolutions the microscopic particle system can be well described, on 
the microscopic scale, by a local version of the equilibrium measure 
which is stationary under the dynamics. These measures depend on 
quantities conserved by the microscopic dynamics, such as the particle
density, which then evolve on the slower hydrodynamic time scale 
according to the hydrodynamic equations. This requires good mixing or 
chaotic properties of the dynamics (as well as of all the relevant 
equilibrium states). This is particularly so for the case of diffusive
scaling where longer times are involved. It is for this reason that
the only model systems of interacting particles for which the HSL has 
been established in the diffusive limit are systems with stochastic 
dynamics. Thus the HSL for Ginzburg-Landau models was established
first by Guo, Papanicolaou and Varadhan, [\rcite{GPV}], by applying 
entropy techniques. These techniques were further developed by
Rezakhanlou, [\rcite{R}], to cover the case when the invariant measure
is not a product measure and phase transitions may occur. These
methods can be applied also to lattice gas models that satisfy the so 
called ``gradient condition'', [\rcite{S}]. For lattice gases this 
condition is however not natural and the only known examples are when 
the invariant measure is a product measure or the spatial dimension is
one, [\rcite{KLS}], [\rcite{S}]. Very recently the diffusive HSL for
non gradient lattice gases has been proved by Varadhan and Yau, 
[\rcite{VY}]. For systems of particles in the continuum the gradient 
condition is more natural, while a common technical problem in these 
models is the control of the local number of particles: the
conservation law cannot prevent locally very high densities. The only 
continuum models treated with the entropy techniques quoted above are 
one dimensional systems of Brownian particles interacting via positive
superstable short range potentials considered by Varadhan,
[\rcite{V}], and Ornstein-Uhlenbeck interacting processes studied by
Olla and Varadhan, [\rcite{OV}]. We should also mention here that the 
diffusive limit can be proven for a Hamiltonian system of
non-interacting particles moving among a fixed array of convex hard 
scatterers: the Sinai billiard system with finite horizon in $d=2$, 
[\rcite{BS}], [\rcite{BCS}], [\rcite{LS1}], [\rcite{LS2}]. 

In 1991 Yau, [\rcite{Y}], proposed a new method for proving the HSL
of interacting particle systems of gradient type, looking at the 
relative entropy and its rate of change w.r.t. local Gibbs states. 
This method can be applied also to continuum systems in higher
dimension, e.g. in the derivation of the Euler equations from a 
Hamiltonian system with weak noise considered by Olla, Varadhan, and 
Yau, [\rcite{OVY}]. 

In the present paper we extend the work of Varadhan to Brownian
particles with positive superstable short range potentials in all
dimensions. In addition we also permit long range pair interactions
of the Kac type in which the range parameter $\g^{-1}$ goes to
infinity as the macro to micro spatial scale $\ve^{-1}$. This extends
previous work for such systems on a lattice, [\rcite{GL1}]. 

To be more precise, we consider a system of $N$ particles which
evolve in time according to the non-inertial Brownian dynamics
   $$
{dr_i\over d\t} = - {\partial\Psi\over\partial r_i}(r_1,\ldots, 
r_N) + {\cal W}_i(\t)
   \Eq(0p1)
   $$
where ${\cal W}_i(\t)$ is a stochastic Langevin force with
Gaussian statistics having covariance $(\beta/2) \d_{ij}\d(\t-\t')
\underline {\bf 1}$, $\underline{\bf 1}$ the unit $d$-dimensional
tensor. The parameter $\b$ is the inverse temperature of the canonical
ensemble, $\mu \sim \exp[-\b\Psi]$, which is the stationary measure
for the evolution. The potential energy $\Psi$ is a sum of pair 
potentials,
   $$
\Psi(r_1,\ldots, r_N) = {1\over 2} \sum_{i\ne j} \big[ V(r_{ij})
+ \g^d J(\g r_{ij})\big]
   \Eq(0p2)
   $$
where $r_{ij}= r_i-r_j$ and the $r_i$, $i=1,\ldots,N$, are confined
to a $d$-dimensional torus ${\cal T}^d_L$ of length $L$. We take 
$V(r)$ to have a finite range $R$ with $R<\g L$: $\g^{-1}$ is the 
range of the Kac potential which will be taken to be large compared 
to the inter-particle spacing $L/N^{1/d}$. Systems with interaction of
form \equ(0p2), with $V(r)\equiv 0$, $J(r)> 0$, and different types
of dynamics, have been investigated numerically and analytically by
Klein and coworkers as model of glassy dynamics, [\rcite{GK}], 
[\rcite{K}], [\rcite{KGRCM}].
 
We observe that due of to the prefactor $\g^{d+1}$ appearing in the 
force term due to the Kac potential, the dynamics defined by \equ(0p1)
is a weak perturbation of the one defined for $J=0$, i.e. without 
long range interactions. Thus we may expect that for small $\g$'s 
the system reaches local equilibrium w.r.t. the short range potential
on spatial scales smaller than $\g^{-1}$ at times of order
$\g^{-2}$. The effect of the long range interaction on such states
will then appear only in determining the macroscopic equation 
for the relevant parameters describing the local equilibrium. 

In fact we shall take as our initial distribution something close
to the local equilibrium distribution relative to the short range
potential $V$ with a density which varies on the scale of 
$L\sim N^{1/d}\sim \g^{-1}$ and consider macroscopic times of order
$\tau /\g^{-2}$. The HSL will then correspond to letting 
$\g\downarrow 0$. We will prove that in that limit the density profile
on the macroscopic scales $x$ and $t$ will satisfy the following 
non-local integro-differential equation of the diffusive type:
   $$
{\partial\rho\over\partial t}(t,x) = \nabla\cdot
\bigg\{D\big(\rho(t,x)\big)\nabla\rho(t,x) + \sigma\big(\rho(t,x)\big) 
\int_{{\cal T}^d}\!dy\,\nabla J(x-y)\rho(t,y) \bigg\}
   \Eq(0p3)
   $$
where the integral is over the $d$-dimensional unit torus ${\cal T}^d$
and $\sigma(\rho)\equiv\b\rho$ is the mobility of a system of interacting
Brownian particles, which, due to the fact that the system is
gradient, does not depend on the interactions, [\rcite{S}]. 
The diffusion coefficient $D(\rho)$ is given explicitly in terms of
the Helmholtz free energy density $a(\b,\rho)$ associated to the 
``reference system'' interacting only with the short range potential 
$V$, in such a way that the following ``Einstein Relation'' holds 
([\rcite{S}]): 
   $$
D(\rho) = \sigma(\rho){\partial\l\over\partial\rho} =
\sigma(\rho) {\partial^2 a \over \partial\rho^2}
   \Eq(0p4)
   $$
where $\l$ is the chemical potential of the reference system at
density $\rho$. As in the lattice case, [\rcite{GL1}], eq. \equ(0p3) 
can be rewritten in terms of the gradient flux associated to the 
classical local mean field free energy functional and the density 
dependent mobility $\sigma(\rho)$:
   $$
{\partial\rho\over\partial t}(t,x) = \nabla\cdot\bigg\{\sigma(\rho)
\nabla {\delta{\cal F}\over\delta\rho}\bigg\}(t,x)
   \Eq(0p5)
   $$
where
   $$
{\cal F}(\rho) = \int_{{\cal T}^d}\!dx\, a(\b,\rho(x)) + 
{1\over 2}\int_{{\cal T}^d}\!dx\int_{{\cal T}^d}\!dy\,
J(x-y)\rho(x)\rho(y)
   \Eq(0p6)
   $$

Our proof is based on Yau's method quoted above. The main restriction
of this method is that the derivation of the HSL is valid only as long
as the macroscopic equation has a smooth classical solution. 
Consequently, unlike the lattice case, we can no longer guarantee 
existence of global solutions. In fact, even if the initial datum is 
smooth and lies in the one phase region (for the reference system), 
we cannot guarantee that the time evolution will not develop
singularities or create regions of high density where the reference
system undergoes a phase transition and the diffusion coefficient 
$D(\rho)$ vanishes. 

The outline of the rest of the paper is as follows. In Section 2 
we give a precise description of our system and present the results.
In Section 3 we prove the HSL by computing the relative entropy and
its rate of change w.r.t. the local equilibrium states of the reference
system. To do this we need a local ergodic theorem whose proof is
sketched in Section 4 and large deviation estimates for the local
Gibbs states which are the content of Section 5. A local existence 
theorem of classical solutions for the macroscopic equation is quite
standard, a sketch of the proof is given at the end of Section 3.

\goodbreak
\vskip1truecm
\centerline {\bf 2. Notation and results.}

\vskip.5truecm
\numsec= 2
\numfor= 1
\numtheo=1

In this section we state our problem in a precise mathematical form
using from the beginning the rescaled space and time variables,
$x_i=\g r_i$, and $t = \g^2 \t$. We also absorb $\beta/2$ into the
Brownian motion term which remains invariant under this rescaling of
space and time. In these units we consider a system of N interacting 
Brownian motions $\underline x(t) = \{x_1(t), \dots, x_N(t)\}$  with 
state space $\TT^d$, the $d$-dimensional unit torus, satisfying the 
following equations ($i=1,\dots, N$):
   $$
dx_i = - \b \bigg[ \g^{-1} \sum_{j: j \ne i} \nabla V
(\g^{-1}(x_i-x_j)) + \g^d \sum_{j:j \ne i} 
\nabla J (x_i-x_j)\bigg] dt + \sqrt{2} dw_i
   \Eq(p1)
   $$
where $\{w_1,\dots,w_N\}$ are independent Brownian motions on $\TT^d$,
the parameter $\b\ge 0$ is the inverse temperature, $J\in C^2(\TT^d)$ 
and $V(r) \in C^1(\R^d)$ is a positive function of $|r|$, with compact
support and such that $V(0)>0$. The latter implies that $V$ is 
superstable. In eq. \equ(p1), $\nabla V(\g^{-1}(x_i-x_j))$ and
$\nabla J(x_i-x_j)$ are the gradients of the functions $V(\cdot)$ and 
$J(\cdot)$ w.r.t. their arguments, evaluated at the points 
$\g^{-1}(x_i-x_j)$ and $(x_i-x_j)$ respectively.
We shall further assume that the number of particles $N$ 
depends on the scaling parameter $\g \in (0,1]$ in such a way that 
$N\g^d \nearrow 1$ as $\g\downarrow 0$ (typically $N= [\g^{-d}]$).

The process $t \to \underline x (t)$ is a diffusion on $\TT^{dN}$ with 
generator 
   $$
L_\g = L_\g^{(0)} + U_\g
   \Eq(p2)
   $$
where
   $$
L_\g^{(0)} = \sum_i \Delta_i - \g^{-1} \sum_{i\ne j}
\b\nabla V (\g^{-1}(x_i-x_j)) \cdot \nabla_i 
   \Eq(p3)
   $$
and 
   $$
U_\g = - \g^d \sum_{i\ne j} \b \nabla J (x_i - x_j) \cdot \nabla_i
   \Eq(p4)
   $$
In \equ(p3) and \equ(p4) $\Delta_i$ ($\nabla_i$) denotes the Laplacian
(gradient) w.r.t. the $i$-th particle component of $\underline x \in 
\TT^{dN}$. Note that the diffusion $L_\g^{(0)}$ is reversible w.r.t. 
   $$
\mu_\g(d\underline x) = {1\over Z_\g} \exp\bigg[-{\b\over 2}
\sum_{i\ne j} V(\g^{-1}(x_i-x_j))\bigg] d\underline x
   \Eq(p5)
   $$
where $Z_\g$ is the normalization factor making $\mu_\g$ a 
probability measure on $\TT^{dN}$.

If the initial distribution of the diffusion has a density 
$f_\g^{(0)}$ w.r.t. $\mu_\g$ then the density at any later time,
$f_\g(t,\underline x)$, satisfies the forward Fokker-Planck 
equation 
   $$
{\partial f_\g \over \partial t} = L_\g^* f_\g, \quad
f_\g\big|_{t=0} = f_\g^{(0)}
   \Eq(p6)
   $$
where $L_\g^*$ is the adjoint of $L_\g$ w.r.t. $\mu_\g$.

To state our result we need to introduce some thermodynamic quantities
relative to the reference system, i.e. the system of particles 
interacting only via the short range (superstable) potential $V$. For 
any regular domain $\L$ of $\R^d$ we define the grand canonical
partition function
   $$ 
Z_\L(\b,\l) = e^{-|\L|} \sum_{N=0}^\infty {e^{\b\l N} \over N!} 
\int\limits_{\L^N} \! dr_1 \cdots dr_N 
\exp\bigg[-{\b\over 2}\sum_{i\ne j}V(r_i-r_j)\bigg]
   \Eq(p7) 
   $$
where $\l\in \R$ is the chemical potential. The pressure is defined
by the limit
   $$
p(\b,\l) = \lim_{\L \nearrow \R^d} {1\over\b |\L|} \log Z_\L(\b,\l)
   \Eq(p7bis)
   $$
which exists and defines a convex and continuous function of $\b$
and $\l$, see [\rcite{R1}] and [\rcite{R2}].

Setting the inverse temperature equal to some fixed value $\b>0$ 
(which we will sometimes omit) there exists, for the reference
system, a non empty open set $\UU \subseteq \R$ such that for any 
$\l \in \UU$ there is a unique (infinite volume) Gibbs state. This is 
a point process on $\R^d$, invariant and ergodic w.r.t. space translations, 
satisfying the DLR equations relative to the potential $V$, see e.g. 
[\rcite{G1}], [\rcite{R1}]. The pressure is a smooth function of 
$\l \in \UU$ and the average density of particles $\rho$, as a function
of the chemical potential, is given by the smooth 1-1 map $\l \mapsto
\rho(\l) = \partial_\l p(\b,\l)$ of $\UU$ onto $\WW \doteq \partial_\l
p(\b,\UU) \subseteq \R_+$. To make more symmetric the correspondence 
between the parameters $\l$ and $\rho$ we introduce the Helmholtz free
energy $a(\b,\rho)$ as the Legendre transform of the pressure:
   $$
a(\b,\rho) \doteq \sup_{\l \in \R}\{\l\rho - p(\b,\l)\}
   \Eq(p8)  
   $$
and we recover the chemical potential as a function of the density by 
the smooth 1-1 map $\rho \mapsto \l(\rho) = \partial_\rho a
(\b,\rho)$ of $\WW$ onto $\UU$. 

We consider the nonlinear non-local integro-differential equation 
\equ(0p3) that we rewrite below in a more concise form:
   $$
{\partial \rho \over \partial t}(t,x) = \nabla \cdot 
\big\{D(\rho)\nabla\rho + \sigma(\rho)\nabla J*\rho\big\}(t,x)
   \Eq(p9)
   $$
where `$*$' denotes convolution on $\TT^d$ and recall that $\sigma(\rho) 
= \b\rho$. In the sequel we will use the capital letter $P$ to denote 
the pressure as a function of the density. Then $P'(\rho) = \rho\l'(\rho)$ 
so that the diffusion coefficient in \equ(p9) is $D(\rho)= \b P'(\rho)$ 
(see \equ(0p4)).

In the one phase region the pressure is a smooth, strictly increasing 
function of the density, so that $D(\rho)$ is smooth and strictly
positive for any $\rho \in \WW$. Then the following theorem holds,
whose proof is sketched at the end of the next section.

\goodbreak
\vskip.2truecm
\noindent{\bf \Theorem (sp1)}
{\sl There exist locally classical solutions of \equ(p9) that lie inside
the one phase region $\WW$.}

\vskip.2truecm
We fix such a solution $\rho(t,x)$, $0\le t\le T$ ($T>0$).
We may assume that there is a compact set $K_w \subset \WW$ such that
$\rho(t,x) \in K_w$ for any $(t,x) \in [0,T]\times\TT^d$ and 
dist$(K_w,\R_+ \setminus\WW) \ge 2\d_1$ for some $\d_1>0$. Clearly
$\l(t,x) \doteq \partial_\rho a(\b,\rho(t,x))$ lies in the compact
set $K_u \doteq \partial_\rho a(\b,K_w)$ and 
dist$(K_u,\R\setminus\UU)\ge 2\d_2$ for some $\d_2>0$.

We introduce the local Gibbs state associated to the above macroscopic 
evolution $\rho(t,x)$ as the probability measure on $\TT^{dN}$ which 
is absolutely continuous w.r.t. $\mu_\g(d\underline x)$ with density
   $$
\hat f_\g(t,\underline x) = {1\over C_\g(t)} \exp\bigg[ \sum_i 
\b\l(t,x_i) \bigg]
   \Eq(p10)
   $$
where $C_\g(t)$ is the normalization constant making $\hat f_\g$ a 
probability density.

Our main result is

\goodbreak
\vskip.2truecm
\noindent{\bf \Theorem (sp2)}
{\sl Let $f_\g$ be the solution of the Fokker-Plank equation \equ(p6)
with an initial distribution $f_\g^{(0)}$ such that 
   $$
\lim_{\g \downarrow 0} \g^d \int \! \mu_\g(d\underline x) 
f_\g^{(0)}(\underline x) \log{f_\g^{(0)}(\underline x) \over 
\hat f_\g(0,\underline x)} = 0
   \Eq(p11)
   $$  
Then, for any $\vf \in C^\infty(\TT^d)$, any $\d>0$, and any $t\in [0,T]$,  
   $$
\lim_{\g \downarrow 0} \int_{A_{\d,\vf}^t} \! \mu_\g(d\underline x) 
f_\g(t,\underline x) = 0
   \Eq(p12)
   $$
where 
   $$
A_{\d,\vf}^t \doteq \bigg\{\underline x \in \TT^{dN} \, : \,
\bigg| N^{-1} \sum_i \vf (x_i) - \int_{\TT^d}\!dx\, \vf (x) 
\rho(t,x)\bigg| > \d \bigg\}
   $$
}
\vskip.2truecm

\noindent {\it Notation:} From now on we will write $f(t,\cdot) =
f(t)$ for functions on $[0,T] \times \TT^d$ or $[0,T] \times \TT^{dN}$.
  
We will prove Theorem \equ(sp2) by using the relative entropy method 
introduced by Yau, [\rcite{Y}]. We recall the {\sl basic entropy
estimate}: if $\mu,\nu$ are two probability measures on the same
measurable space, then for any $F\in L^1(d\nu)$,
   $$
\int\! d\mu\, F \le H(\mu|\nu) + \log\int\! d\nu \, \exp[F] 
   $$
where $H(\mu|\nu)$ is the relative entropy of $\mu$ w.r.t. $\nu$ and,
if $\mu\ll\nu$,
   $$
H(\mu|\nu) = \int\!d\mu\,\log{d\mu\over d\nu}
   $$
For any $t \in [0,T]$ define the functional
   $$
H_\g(t) \doteq \g^d \int \! \mu_\g(d\underline x) f_\g(t,\underline x)
\log{f_\g(t,\underline x)\over \hat f_\g(t,\underline x)}
   \Eq(p13)
   $$
Note that $H_\g(t)$ is $\g^d$ times the relative entropy of 
$f_\g(t)d\mu_\g$ w.r.t. $\hat f_\g(t)d\mu_\g$ and that the 
argument of the limit in the l.h.s. of \equ(p11) is exactly
$H_\g(0)$. In the next section we will prove:

\goodbreak
\vskip.2truecm
\noindent{\bf \Theorem (sp3)}
{\sl Under the same hypothesis of Theorem \equ(sp2), for any $t \in [0,T]$,
   $$
\lim_{\g \downarrow 0} H_\g(t) = 0 
   \Eq(p14)
   $$
}
\vskip.2truecm

The hydrodynamic limit \equ(p12) follows as a corollary of Theorem 
\equ(sp3). To see this we note that it follows from the large
deviation principle (LDP) for the local Gibbs states \equ(p10), 
see Section 5, that there is a $c(\d,\vf)>0$ such that
   $$
\E^{\hat f_\g(t)}\big[\1_{A_{\d,\vf}^t}\big] \le \exp\big[-c(\d,\vf)N]
   $$
where $\E^f[\cdot]$ denotes the expectation w.r.t. the measure
$fd\mu_\g$ and $\1_\Gamma$ is the characteristic function of the set
$\Gamma$. On the other hand, from the basic entropy estimate the 
following inequality holds (see e.g. [\rcite{Y}]):  
   $$
\E^{f_\g(t)}\big[\1_{A_{\d,\vf}^t}\big] \le {\log 2 + \g^{-d}H_\g(t)
\over  \log\Big(1 + \E^{\hat f_\g(t)}\big[\1_{A_{\d,\vf}^t}\big]^{-1}
\Big)}
   $$
so that, for some $C>0$, $\E^{f_\g(t)}\big[\1_{A_{\d,\vf}^t}\big] \le 
C \big( N^{-1} + (N\g^d)^{-1}H_\g(t)\big) \to 0$ as $\g \downarrow 0$.

\goodbreak
\vskip1truecm
\centerline {\bf 3. Proof of Theorems \equ(sp3) and \equ(sp1).}

\vskip.5truecm
\numsec= 3
\numfor= 1
\numtheo=1

Because of the hypothesis \equ(p11) on the initial distribution,
we only need a good estimate on the time derivative of $H_\g(t)$. 
By Lemma 3.1 of [\rcite{OVY}], the following bound holds:
   $$
{dH_\g\over dt} \le \g^d \int\!\mu_\g(d\underline x) f_\g(t,
\underline x) \hat f_\g(t,\underline x)^{-1} \bigg( L_\g^*
- {\partial \over \partial t} \bigg) \hat f_\g(t,\underline x)
   \Eq(p15)
   $$
Recalling \equ(p2) and that $L_\g^{(0)}$ is reversible w.r.t. 
$\mu_\g(d\underline x)$, \equ(p15) gives
   $$
\eqalign{
{dH_\g\over dt} & \le \g^d \int\!\mu_\g(d\underline x) f_\g(t,
\underline x) \hat f_\g(t,\underline x)^{-1} \bigg( L_\g^{(0)}
- {\partial \over \partial t} \bigg) \hat f_\g(t,\underline x) 
\cr & + \g^d \int\!\mu_\g(d\underline x) f_\g(t,\underline x) 
\hat f_\g(t,\underline x)^{-1} U_\g^* \hat f_\g(t,\underline x) 
}
   \Eq(p16)
   $$
where $U_\g$ is defined in \equ(p4). By an explicit computation
   $$
\eqalign{
\hat f_\g(t,\underline x)^{-1} \bigg( L_\g^{(0)} & - 
{\partial \over \partial t} \bigg) \hat f_\g(t,\underline x) 
= \b \sum_i \bigg\{\b|\nabla\l|^2(t,x_i) + \Delta\l(t,x_i)  - 
\dot\l(t,x_i) \cr & - \g^{-1} 
\sum_{j:j\ne i} \b\nabla V(\g^{-1}(x_i-x_j)) \cdot \nabla\l(t,x_i)
\bigg\} + \E^{\hat f_\g(t)}\bigg[ \sum_i \b \dot\l (t,x_i)\bigg]
}
   \Eq(p17)
   $$
where $\dot \l$ denotes the time derivative of $\l$. 
Integrating by parts one computes the action of the adjoint 
operator $U_\g^*$ and gets
   $$
\eqalign{
\hat f_\g(t,\underline x)^{-1} U_\g^* \hat f_\g(t,\underline x)
& = \g^d \b \sum_{i\ne j} \bigg\{ \b \nabla J (x_i-x_j)\cdot 
\nabla\l(t,x_i) + \Delta J(x_i-x_j) \cr & - \g^{-1} \sum_{k:k \ne i}
\b \nabla V (\g^{-1}(x_i-x_k))\cdot\nabla J(x_i - x_j) \bigg\}
}
   \Eq(p18)
   $$
In both \equ(p17) and \equ(p18) there is a term of the following type:
   $$
K_V(\underline x) = \sum_{i\ne k} \g^{-1} \nabla V(\g^{-1}(x_i-x_k))
\cdot \nabla\vf (x_i)
   $$
for some smooth function $\vf$ on $\TT^d$ (actually $\vf=\b^2\l(t,\cdot)$
and $\vf=\b^2 J(\cdot - x_j)$ in \equ(p17) and \equ(p18) respectively). 
Since $\nabla V$ is an odd function, 
   $$
\eqalign{
K_V(\underline x) & = {1\over 2} \sum_{i\ne k} \g^{-1} \nabla V
(\g^{-1}(x_i-x_k))\cdot \big(\nabla \vf (x_i) - \nabla \vf (x_k)\big)
\cr & = - {1\over 2} \sum_{i\ne k} \sum_{\xi,\eta} (\nabla V)^\xi 
(\g^{-1}(x_i - x_k)) D_{\xi\eta}\vf (x_i) 
(\g^{-1}(x_i-x_k))^\eta + R_V(\underline x)
}
   \Eq(p19)
   $$
where $x^\xi$ is the $\xi$-th component of $x \in \TT^d$ and 
$D_{\xi\eta} = \partial^2/(\partial x^\xi \partial x^\eta)$.
Since $\nabla V$ has compact support, we can estimate the 
reminder using Taylor expansion:
   $$
|R_V(\underline x)| \le r_\g(\vf) \sum_{i\ne k} |\nabla V|
(\g^{-1}(x_i -x_k))
   $$
with $r_\g(\vf)\to 0$ as $\g \downarrow 0$. 

Inserting \equ(p17) and \equ(p18) into \equ(p16) and using \equ(p19)
we get
   $$
{dH_\g\over dt}  \le \E^{f_\g(t)}\big[\b\Phi_\g(\underline x,\l(t)) 
\big] + \E^{\hat f_\g(t)}\bigg[ \sum_i \b\dot\l (t,x_i)\bigg] +
\varepsilon_\g(t)
   \Eq(p20)
   $$
where
   $$
\eqalign{
\Phi_\g(\underline x, \l(t)) & = \g^d \sum_i \bigg\{
\b |\nabla\l|^2(t,x_i) + \Delta\l(t,x_i) - \dot\l(t,x_i)
\bigg\} \cr & + \g^d \sum_{i\ne j}  \sum_{\xi,\eta} {\b\over 2} 
(\nabla V)^\xi (\g^{-1}(x_i - x_j)) D_{\xi\eta} \l (t,x_i) 
(\g^{-1}(x_i-x_j))^\eta \cr & + \g^{2d} \sum_{i\ne j}
\big\{\b\nabla \l (t,x_i) \cdot \nabla J (x_i-x_j) + \Delta J (x_i-x_j)
\big\} \cr & + \g^{2d} \sum_{i\ne j} \sum_{k:k\ne i} \sum_{\xi,\eta}
{\b\over 2} (\nabla V)^\xi (\g^{-1}(x_i - x_k))
D_{\xi\eta}J(x_i-x_j)(\g^{-1}(x_i-x_k))^\eta 
}
   \Eq(p21)
   $$
while $\varepsilon_\g(t)$ satisfies the bound
   $$
|\varepsilon_\g(t)| \le \g^d \big( r_\g(\b^2\l(t)) + N\g^d 
r_\g(\b^2 J) \big) \E^{f_\g(t)}\bigg[ \sum_{i\ne j}
|\nabla V|(\g^{-1}(x_i-x_j)) \bigg]
   $$
To obtain the behavior of $\varepsilon_\g(t)$ when $\g \downarrow 0$
we use the following lemma which is proved in Section 4:

\goodbreak
\vskip.2truecm
\noindent{\bf \Lemma (sp4)}
{\sl If $W$ is a continuous function on $\R^d$ with compact support,
there is $C_W>0$ such that, for any $\g\in (0,1]$ and any $t\in (0,T]$,
   $$
\E^{f_\g(t)}\bigg[\g^d \sum_{i\ne j} 
W(\g^{-1}(x_i-x_j)) \bigg] \le C_W 
   $$
}
\vskip.2truecm
By applying the lemma with $W=|\nabla V|$ we conclude that
   $$
\lim_{\g\downarrow 0} \sup_{t\in [0,T]}|\varepsilon_\g(t)| = 0
   \Eq(p22)
   $$
Moreover, by the LDP for the local Gibbs state, see Section 5,
for any $t\in [0,T]$,
   $$
\lim_{\g\downarrow 0} \E^{\hat f_\g(t)}\bigg[ \sum_i \dot\l
(t,x_i)\bigg] = \int_{\TT^d}\! dx\, \dot \l (t,x) \rho(t,x)
   \Eq(p23)
   $$

Now we want to write $\Phi_\g(\underline x,\l(t))$ in terms of local 
empirical quantities. Let $\O$ be the space of particle configurations 
on $\R^d$, i.e. $\o \in \O$ is a subset of $\R^d$ which is locally 
finite (see Section 5 for more details). Given $x\in \TT^d$, for any
$\underline x \in \TT^{dN}$ we construct a configuration $\o_{x,\g}
\in \O$ by setting
   $$
\o_{x,\g} \doteq \big\{ q_i = \g^{-1}(x_i-x) \, :\, |x_i-x|<1/4 \big\}
   $$
(since there is no risk of confusion, to simplify notation we omit 
the explicit dependence on $\underline x$ of $\o_{x,\g}$). 
Clearly $\o_{x,\g}$ is well defined in every compact set inside 
the cube of $\R^d$ of side $1/(2\g)$ and centered in the origin. So, if
$F(\o)$ is a local function on $\O$, $F(\o_{x,\g})$ is well
defined for any $\g$ small enough.

Let us introduce the cubes $D_n = \{q\in\R^d\, : \, |q^\xi|\le n, \,\,
\xi=1,\ldots ,d\}$, $n \in \N$. For any local function $F$ we denote by
$F_n$ its spatial average over the cube $D_n$, i.e.
   $$
F_n(\o) \doteq {1\over |D_n|} \int_{D_n} \! dr \, F(\t_r\o)
   \Eq(p23bis)
   $$
where $\t_r$ is the space translation by $r$ ($\t_r r'=r+r'$). 

Let $\chi$ be any non negative function on $\R^d$ with compact
support and total integral 1. We define the following local
functions on $\O$:
   $$
R(\o) \doteq \sum_{q\in\o} \chi(q), \Eq(p24)
   $$
   $$
G^{\xi\eta}(\o) \doteq {\b\over 2} \sum_{\scriptstyle q,q' \in \o 
\atop \scriptstyle q \ne q'} \chi (q) (\nabla V)^\xi(q-q')(q-q')^\eta, 
\quad \xi,\eta =1,\dots ,d
   \Eq(p25)
   $$
and let $R_n(\o)$, $G^{\xi\eta}_n(\o)$ be their averages over $D_n$.
Observe that $R(\o)$ is a natural version of local density for the
configuration $\o$, while $G^{\xi\eta}(\o)$ is the local quantity 
appearing in the virial theorem (see below, before \equ(p32)).

\goodbreak
\vskip.2truecm
\noindent{\bf \Lemma (sp5)}
{\sl Let $\vf,\psi$ be smooth functions on $\TT^d$. Then
   $$
\limsup_{n\to\infty}\limsup_{\g\downarrow 0} \sup_{t\in [0,T]}
\E^{f_\g(t)} \bigg| \g^d \sum_i \vf(x_i) - \int_{\TT^d}
\! dx\, \vf (x) R_n(\o_{x,\g}) \bigg| = 0 
   \Eq(p26)
   $$
   $$
\eqalign{
\limsup_{n\to\infty}\limsup_{\g\downarrow 0} \sup_{t\in [0,T]}
& \E^{f_\g(t)} \bigg| \g^d \sum_{i\ne j} {\b\over 2}(\nabla V)^\xi
(\g^{-1}(x_i-x_j))(\g^{-1}(x_i-x_j))^\eta \vf(x_i) \cr & - 
\int_{\TT^d}\! dx\, \vf (x) G^{\xi\eta}_n(\o_{x,\g}) \bigg| = 0 
} \Eq(p27)
   $$
   $$
\eqalign{
\limsup_{n\to\infty}\limsup_{\g\downarrow 0} \sup_{t\in [0,T]}
& \E^{f_\g(t)} \bigg| \g^{2d} \sum_{i\ne j} \vf(x_i) \psi(x_i-x_j)
\cr & -\int_{\TT^d} \! dx \int_{\TT^d} \! dy \, \vf (x) R_n(\o_{x,\g}) 
\psi(x-y) R_n(\o_{y,\g}) \bigg| = 0 
} \Eq(p28)
   $$
   $$
\eqalign{
\limsup_{n\to\infty}\limsup_{\g\downarrow 0} \sup_{t\in [0,T]}
& \E^{f_\g(t)} \bigg| \g^{2d} \sum_{i\ne j} \sum_{k:k\ne i}{\b\over 2} 
(\nabla V)^\xi (\g^{-1}(x_i-x_k))(\g^{-1}(x_i-x_k))^\eta \cr &
\times \psi(x_i-x_j) -\int_{\TT^d} \! dx \int_{\TT^d} \! dy \, 
G^{\xi\eta}_n(\o_{x,\g}) \psi(x-y) R_n(\o_{y,\g}) \bigg|  = 0 
} \Eq(p29)
   $$
}
\vskip.2truecm

\noindent {\it Proof.} We have
   $$
\int_{\TT^d} \! dx\, \vf (x) R_n(\o_{x,\g}) = 
\g^d \sum_i \vf_{n,\g}(x_i)
   $$
where, for $z\in \TT^d$, 
   $$
\vf_{n,\g}(z) \doteq \g^{-d} \int_{\TT^d} \! dx\, \vf (x) {1\over |D_n|}  
\int_{D_n} \! dq \, \chi (\g^{-1}(z-x) + q) 
   $$
Then the expectation in the l.h.s. of \equ(p26) can be bounded by 
$\|\vf - \vf_{n,\g}\|_\infty$ that vanishes as $\g\downarrow 0$ for  
any $n \in \N$ because of the smoothness assumptions on $\vf$. In an 
analogous way one can estimate the expectation in the l.h.s. of 
\equ(p27) by
   $$
\|\vf - \vf_{n,\g}\|_\infty \g^d  \E^{f_\g(t)}\bigg[ \sum_{i\ne j}
{\b\over 2} |\nabla V|(\g^{-1}(x_i-x_j))\big|\g^{-1}(x_i-x_j)\big| 
\bigg]
   $$
and \equ(p27) follows from Lemma \equ(sp4). 

Let us consider now \equ(p28). We have
   $$
\int_{\TT^d} \! dx \int_{\TT^d} \! dy \, \vf (x) R_n(\o_{x,\g}) 
\psi(x-y) R_n(\o_{y,\g}) = \g^{2d} \sum_{i\ne j} \vf_{n,\g}(x_i)
\psi_{n,\g}(x_i - x_j) + Err_1
   $$
where $\psi_{n,\g}$ is defined as $\vf_{n,\g}$ and 
   $$
\big|Err_1\big| \le \|\vf_{n,\g}\|_\infty \max\{
|\psi_{n,\g}(x-z) - \psi_{n,\g}(y-z)| \, : \, x,y,z \in \TT^d, |x-y|
\le \g (r + n) \}
   $$
with $r$ such that the support of $\chi$ is contained in the closed
ball of radius $r$. Since $|\psi_{n,\g}(x-z) - \psi_{n,\g}(y-z)| \le
\|\nabla\psi\|_\infty\g (r+n) + 2 \|\psi - \psi_{n,\g}\|_\infty$,
$Err_1$ vanishes as $\g \downarrow 0$. Since $|\vf_{n,\g}(x_i)
\psi_{n,\g}(x_i - x_j) - \vf (x_i) \psi (x_i - x_j)| \le
\|\vf\|_\infty \|\psi - \psi_{n,\g}\|_\infty + 
\|\psi\|_\infty \|\vf - \vf_{n,\g}\|_\infty$, \equ(p28) follows.
In the same manner we compute
   $$
\eqalign{
\int_{\TT^d} & \! dx \int_{\TT^d} \! dy \, G^{\xi\eta}_n(\o_{x,\g}) 
\psi(x-y) R_n(\o_{y,\g}) \cr & = \g^{2d} \sum_{i\ne j} \psi_{n,\g}
(x_i - x_j) \sum_{k:k\ne i} {\b\over 2} (\nabla V)^\xi (\g^{-1}(x_i - x_k))
(\g^{-1}(x_i-x_k))^\eta + Err_2
}
   $$
with 
   $$
\eqalign{
|Err_2| & \le \max\{|\psi_{n,\g}(x-z) - \psi_{n,\g}(y-z)|
\, : \, x,y,z \in \TT^d, |x-y| \le \g (r + n) \} \cr &
\times \g^d \sum_{i\ne j}|\nabla V|(\g^{-1}(x_i - x_j))
\big|\g^{-1}(x_i-x_j)\big|
}
   $$
Then \equ(p29) follows from Lemma \equ(sp4). \qed

\vskip.2truecm

Collecting together \equ(p20), \equ(p22), \equ(p23) and applying Lemma
\equ(sp5) to $\E^{f_\g(t)}\big[ \Phi_\g(\underline x, \l(t))\big]$, we 
obtain
   $$
\limsup_{n\to\infty}\limsup_{\g\downarrow 0} \sup_{t\in [0,T]}
\bigg\{ {dH_\g\over dt} + \E^{f_\g(t)} \bigg[\b \int_{\TT^d}\! dx\, 
\big( \Psi_{\underline x}(t,x) - \dot \l (t,x)\rho (t,x)\big)\bigg] 
\bigg\} \le 0 
   \Eq(p30)
   $$
with
   $$
\eqalign{
\Psi_{\underline x}(t,x) & = 
\bigg[\dot\l (t,x) - \b |\nabla\l|^2(t,x) - \Delta\l (t,x) \bigg] 
R_n(\o_{x,\g}) - \sum_{\xi,\eta} D_{\xi\eta}\l (t,x) G^{\xi\eta}_n(\o_{x,\g})
\cr & - \b\nabla\l (t,x) \cdot \big(\nabla J * R_n(\o_{\cdot,\g})\big)
(x) R_n(\o_{x,\g}) - \big(\Delta J * R_n(\o_{\cdot,\g})\big)(x) 
R_n(\o_{x,\g}) \cr & - \sum_{\xi,\eta} 
\big(D_{\xi\eta} J * R_n(\o_{\cdot,\g})\big)(x) G^{\xi\eta}_n(\o_{x,\g})
}
   \Eq(p31)
   $$
where, for any $\vf \in C(\TT^d)$,
   $$
\big(\vf * R_n(\o_{\cdot,\g})\big)(x) \doteq \int_{\TT^d}\! dx\,
\vf(x-y) R_n(\o_{y,\g})
   $$

Now we want to substitute the spatial average $G^{\xi\eta}_n(\o_{x,\g})$
with a function of the empirical density $R_n(\o_{x,\g})$. More 
precisely we would like to replace it by the average of $G^{\xi\eta}$ 
w.r.t. the Gibbs state with density equal to $R_n(\o_{x,\g})$. 

To do this we need to introduce some cutoffs. Let $K$ be a compact
set such that 
   $$
K_w \subset K \subset \WW, \quad {\rm dist}(K,\R_+\setminus \WW) 
\ge \d_1, \quad {\rm dist}(K_w,\R_+\setminus K) \ge \d_1
   $$
(recall that $K_w$ is the compact set inside the one phase region
$\WW$ where the solution $\rho (t,x)$ lies and that dist$(K_w,
\R_+\setminus \WW)\ge 2\d_1$) and define the local
function $u_n(\o) \doteq \1_K(R_n(\o))$. We denote also by $\phi_k$ 
the cutoff at the level $k \in \R_+$, i.e. $\phi_k(s) = s$ if $|s| 
\le k$, $\phi_k(s) = \sign(s)\, k$ otherwise.
Finally let $\hat G^{\xi\eta}(\rho)$, $\rho \in \WW$, be the average of
$G^{\xi\eta}(\o)$ w.r.t. the unique Gibbs measure with density
$\rho$. By the virial theorem, see e.g. [\rcite{V}],
   $$
\hat G^{\xi\eta}(\rho) = \big( \b P(\rho) - \rho\big) \d_{\xi\eta}
   $$
where $P(\rho)$ is the pressure as a function of the density $\rho$
introduced just after \equ(p9).

For any measurable function $m:\TT^d\to\R_+$ we define the functional
   $$
\eqalign{
\O(t,x,m) & \doteq \bigg(\dot\l (t,x) - \b |\nabla\l |^2(t,x)
\bigg) m(x) - \b \Delta \l(t,x) P(m(x)) \cr & - \b\nabla\l(t,x) 
\cdot (\nabla J*m)(x)m(x) - \b (\Delta J*m)(x)P(m(x))
}
   \Eq(p32)
   $$
Observe now that, since $P'(\rho)=\rho\l'(\rho)$, 
by integration by parts,
   $$
\eqalign{
\int_{\TT^d}\! dx\, P(\rho(t,x))\Delta\l (t,x) & = 
- \int_{\TT^d}\! dx\, \rho(t,x) \l'(\rho(t,x))\nabla\rho(t,x)
\cdot \nabla\l (t,x) \cr & = - \int_{\TT^d}\! dx\,\rho (t,x) 
|\nabla\l (t,x)|^2
}
   $$
and, analogously,
   $$
\int_{\TT^d}\! dx\, P(\rho(t,x))(\Delta J*\rho(t))(x) = 
- \int_{\TT^d}\! dx\, \rho(t,x)\nabla\l (t,x) \cdot
(\nabla J*\rho(t))(x)
   $$
so that, for any $t\in [0,T]$,
   $$
\int_{\TT^d}\! dx\, \dot\l (t,x) \rho (t,x) = 
\int_{\TT^d}\! dx\, \O (t,x,\rho(t))
   $$
Then we can replace $\dot\l (t,x) \rho (t,x)$ by $\O (t,x,\rho(t))$ 
in \equ(p30). 

We decompose now, for any $k>0$,
   $$
\Psi_{\underline x}(t,x) - \O (t,x,\rho(t)) = \sum_{p=1}^4 \O_p(t,x)
   $$
with
   $$
\eqalign{&
\O_1(t,x) = \big[\O(t,x,R_n(\o_{\cdot,\g})) - \O (t,x,\rho(t)) \big]
u_n(\o_{x,\g}) \cr &
\O_2(t,x) = \big[ \Psi_{\underline x}^{(k)}(t,x) - 
\O(t,x,R_n(\o_{\cdot,\g}))\big] u_n(\o_{x,\g}) \cr &
\O_3(t,x) = \big[ \Psi_{\underline x}^{(k)}(t,x) - 
\O(t,x,\rho(t))\big] (1-u_n(\o_{x,\g})) \cr &
\O_4(t,x) = \big[ \Psi_{\underline x}(t,x) - 
\Psi_{\underline x}^{(k)}(t,x) \big]
}
   \Eq(p33)
   $$
where $\Psi_{\underline x}^{(k)}(t,x)$ is defined as 
$\Psi_{\underline x}(t,x)$ in \equ(p31) with $G^{\xi\eta}_n$
replaced by $(\phi_k \circ G^{\xi\eta})_n$.

In Section 5 we will prove that there is $\d_0>0$ such that 
   $$
\limsup_{k\to\infty}\limsup_{n\to\infty}\limsup_{\g\downarrow 0}
\sup_{t\in [0,T]} \bigg\{\E^{f_\g(t)}\bigg[\int_{\TT^d}\! dx\,
\b\O_p(t,x)\bigg] - \d_0^{-1} H_\g(t)\bigg\} \le 0, \quad p=3,4
   \Eq(p34)
   $$
On the other hand, the local ergodic theorem, see Section 4, implies
that
   $$
\limsup_{k\to\infty}\limsup_{n\to\infty}\limsup_{\g\downarrow 0}
\int_0^T \! ds \, \E^{f_\g(s)}\bigg[\int_{\TT^d}\! dx\, \b
|\O_2(s,x)| \bigg] = 0
   \Eq(p35)
   $$
From \equ(p30), \equ(p34) and \equ(p35) we get, for any $t\in [0,T]$,
   $$
H_\g (t) + \int_0^t \!ds \bigg\{ \E^{f_\g(s)}\bigg[\int_{\TT^d}
\! dx\, \b\O_1(s,x) \bigg] - 2 \d_0^{-1} H_\g (s) \bigg\} \le o(n,\g)
   \Eq(p36)
   $$
with 
   $$
\limsup_{n\to \infty}\limsup_{\g\downarrow 0}o(n,\g) = 0 
   $$
Now, from the basic entropy estimate, for any $\d>0$ and any $s \in
[0,T]$,
   $$
\E^{f_\g(s)}\bigg[\int_{\TT^d} \! dx\, \b\O_1(s,x) \bigg] \ge
- \d^{-1} H_\g(s) - \d^{-1} \g^d \log \E^{\hat f_\g(s)}
\exp \bigg[ - \d\g^{-d} \int_{\TT^d} \! dx\, \b\O_1(s,x) \bigg]
   $$
so that, from \equ(p36), for any $t\in [0,T]$,
   $$
\eqalign{
H_\g(t) & - \d^{-1}\g^d \int_0^t\! ds\, \log \E^{\hat f_\g(s)}
\exp \bigg[ - \d\g^{-d} \int_{\TT^d} \! dx\, \b\O_1(s,x) \bigg]\cr &
- (2\d_0^{-1} + \d^{-1}) \int_0^t \! ds\, H_\g(s) \le o(n,\g)
}
   $$
By applying the Gronwall Lemma to the last inequality we get
   $$
H_\g(t) \le e^{(2\d_0^{-1} + \d^{-1})t} \bigg( o(n,\g) + \d^{-1}\g^d
\int_0^t \! ds \, \log\E^{\hat f_\g (s)}\exp\bigg[-\d\g^d \int_{\TT^d}
\!dx \, \b\O_1(s,x) \bigg]\bigg)
   \Eq(p37)
   $$
In Section 5 we will prove that
   $$
\limsup_{n\to\infty}\limsup_{\g\downarrow 0} \d^{-1}\g^d\log\E^{\hat
f_\g(s)}\exp\bigg[-\d\g^{-d}\int_{\TT^d}\! dx\, \b\O_1(s,x)\bigg] \le
\d^{-1} \Theta_\d(s,\l)
   \Eq(p38)
   $$
where
   $$
\eqalign{
\Theta_\d(s,\l) \doteq \sup\bigg\{& \int_{\TT^d}\! dx\, \Big[\d\b
\big[\O(s,x,m) - \O (s,x,\rho(s)) \big]
\1_K(m(x)) - I_\b(\l(s,x),m(x)) \Big]; \, \cr & m:\TT^d \to 
\R_+ \,\,\,\,\, {\rm integrable} \bigg\}
}
   \Eq(p39)
   $$
and (recall definitions \equ(p7bis) and \equ(p8))
   $$
I_\b(\l,m) \doteq \b \big( p(\b,\l) + a(\b,m) - \l m \big) 
   \Eq(p39bis)
   $$
From \equ(p37) and \equ(p38), for any $t\in [0,T]$,
   $$
\limsup_{\g\downarrow 0} H_\g(t) \le e^{(2\d_0^{-1} + \d^{-1})t}
\int_0^t \! ds \, \d^{-1} \Theta_\d(s,\l)
   \Eq(p40)
   $$
We conclude the proof of Theorem \equ(sp3) from \equ(p40) by showing
that, for $\d$ small enough, $\Theta_\d(s,\l) = 0$ for any $s \in [0,T]$.
We first note that, for $\l \in \UU$, $m \mapsto I_\b(\l,m)$ is strictly
convex on $K$, non negative, and equal to 0 iff $m=\partial_\l p(\b,\l)$, 
so that
   $$
- \int_{\TT^d}\! dx\, I_\b(\l(s,x),m(x)) \le 0, \quad = 0 \,\,\,\,
{\rm iff} \,\,\,\, m(x) 
= \rho(s,x)
   $$
On the other hand the functional
   $$
\GG_s (m) \doteq \int_{\TT^d}\! dx\, \b \big[\O(s,x,m) - \O
(s,x,\rho(s)) \big] \1_K(m(x)) 
   $$
is bounded on the class of functions considered in \equ(p39) and equal 
to 0 for $m=\rho(s)$. Then, for $\d$ small enough, $\Theta_\d(s,\l) = 0$
provided that
   $$
{\d\GG_s\over \d m} \big(\rho(s)\big) = 0
   $$
(observe that $\rho(s)$ is away from $\R_+\setminus K$ because 
dist$(K_w,\R_+\setminus K)\ge \d_1$).
By an explicit computation, 
   $$
\eqalign{
\b^{-1} {\d\GG_s\over \d m} \big(\rho(s)\big) & = \dot\l (s) - 
\b\Big( |\nabla\l|^2(s) + \Delta\l (s) P'(\rho(s)) -
\nabla\l (s) \cdot \nabla J * \rho(s) \cr & - 
\nabla J * \big(\rho(s)\nabla\l (s)\big) - 
P'(\rho(s))\Delta J * \rho(s)  -
\Delta J * P(\rho(s)) \Big)
}
   $$
But, recalling that $P'(\rho) = \rho \l'(\rho) = \b^{-1}D(\rho)$,
   $$
|\nabla\l|^2 + \Delta\l P'(\rho) = \l'(\rho)\big(\nabla\rho\cdot
\nabla\l +\rho\Delta\l\big) = \b^{-1} \l'(\rho) \nabla \cdot 
\big(D(\rho)\nabla\rho\big)
   $$
and
   $$
\eqalign{
\nabla\l \cdot \nabla J * \rho & + \nabla J * (\rho\nabla\l) + 
(\Delta J *\rho) P'(\rho) + (\Delta J * P(\rho)) \cr & = \l'(\rho) 
\big(\nabla\rho \cdot \nabla J * \rho + \rho \Delta J * \rho\big) 
= \l'(\rho) \nabla \cdot \big(\rho\nabla J *\rho\big)
}
   $$
So that, for any $(s,x) \in [0,T]\times \TT^d$, 
   $$
{\d\GG_s\over \d m(x)} \big(\rho(s)\big) = \b\l'(\rho(s,x))\bigg[
{\partial\rho \over \partial s} - \nabla\cdot\big\{D(\rho)\nabla
\rho + \b\rho\nabla J*\rho\big\}\bigg](s,x) = 0
   $$
since $\rho(s,x)$ satisfies \equ(p9) and $\sigma(\rho)=\b\rho$. \qed

\vskip.2truecm
We conclude the section with the proof of Theorem \equ(sp1).
\vskip.2truecm

\noindent {\it Proof of Theorem \equ(sp1) (sketch).} Let $\rho_0\in 
C^2({\cal T}^d)$ be such that $\rho_0({\cal T}^d)\subset\WW$.
By continuity and compactness we can find three compact intervals
$I_i=[a_i,b_i]$, $i=1,2,3$, such that $\rho_0({\cal T}^d)\subseteq
I_1 \subset I_2 \subset I_3 \subset \WW$, $a_3<a_2<a_1<b_1<b_2<b_3$. 
We construct two functions $\tilde D, \tilde\sigma
\in C^1(\R)$ with the following properties: $\tilde D(u) = D(u)$ and
$\tilde\sigma (u) = \sigma (u)$ for $u\in I_2$, $c^{-1}\le \tilde D(u)
\le c$ for some $c>1$ and for any $u\in\R$, supp$(\tilde\sigma) 
\subseteq I_3$. Then we consider the Cauchy problem
   $$
{\partial \rho \over \partial t}(t,x) = \nabla \cdot 
\big\{\tilde D(\rho)\nabla\rho + 
\tilde\sigma(\rho)\nabla J*\rho\big\}(t,x)
   \Eq(p9bis)
   $$
with initial datum $\rho_0$. Arguing exactly as in Theorem 4.1 and 
Remark 4.1 of [\rcite{GL2}], we know that there exists a (unique) 
classical solution $\rho(t,x)$ of the Cauchy problem above (moreover 
it lies in the region $I_3$ at any time).
Clearly we can find $T>0$ such that $\rho (t,x) \in I_2$ for any 
$t \in [0,T]$ and $x\in {\cal T}^d$. From the choice of
$\tilde D$ and $\tilde\sigma$ it follows that $\{\rho(t,x);(t,x) \in
[0,T]\times{\cal T}^d\}$ is also a (local) classical solution of the 
original equation \equ(p9). \qed

\goodbreak
\vskip1truecm
\centerline {\bf 4. Local ergodicity and entropy bounds.}

\vskip.5truecm
\numsec= 4
\numfor= 1
\numtheo=1

We start this section by proving \equ(p35). Since $D_{\xi\eta}\l$ and 
$D_{\xi\eta}J*R_n(\o_{\cdot,\g})$ are bounded functions on $\TT^d$, 
it is sufficient to prove that 
   $$
\eqalign{
\limsup_{k\to\infty}\limsup_{n\to\infty}\limsup_{\g\downarrow 0} &
\int_0^T\! dt\, \E^{f_\g(t)}\bigg[\int_{\TT^d}\! dx\,\big|
(\phi_k \circ G^{\xi\eta})_n(\o_{x,\g}) \cr &
- \hat G^{\xi\eta}(R_n(\o_{x,\g})) \big| u_n(\o_{x,\g})\bigg] = 0
}
   \Eq(p41)
   $$
The main step in proving \equ(p41) is a local ergodic theorem for the
measure $f_\g(t)d\mu_\g$:

\goodbreak
\vskip.2truecm
\noindent{\bf \Theorem (sp6)} {\sl For any local, bounded and 
continuous function $F$,
   $$
\limsup_{n\to\infty}\limsup_{\g\downarrow 0}
\int_0^T\! dt\, \E^{f_\g(t)}\bigg[\int_{\TT^d}\! dx\, \big|
F_n(\o_{x,\g}) - \hat F(R_n(\o_{x,\g}))
\big| u_n(\o_{x,\g})\bigg] = 0
   \Eq(p42)
   $$
where $\hat F(\rho)$, $\rho\in\WW$, is the average of $F$ w.r.t. the 
(unique) Gibbs measure with density $\rho$.
}
\vskip.2truecm

\noindent {\it Proof.} We introduce the translation invariant density
   $$
\bar f_\g (\underline x) = {1\over T} \int_0^T \!ds \, \int_{\TT^d}
\! dx \, f_\g(s,\t_x\underline x)
   \Eq(p43)
   $$
so that \equ(p42) is equivalent to proving that
   $$
\limsup_{n\to\infty}\limsup_{\g\downarrow 0}
\E^{\bar f_\g}\Big[\big| F_n(\o_{0,\g}) - \hat F(R_n(\o_{0,\g}))
\big| u_n(\o_{0,\g})\Big] = 0
   \Eq(p44)
   $$
Let $n_F\in\N $ be such that supp$(F) \subset D_{n_F}$ ($D_n$ as 
in definition \equ(p23bis)) and let $\bar n = n + n_F$. 
Through the mapping $\underline x \mapsto \o_{0,\g}$, for 
any $\g$ small enough it is well defined the projection
   $$
\Pi_{\bar n} : \TT^{dN} \to \O\big|_{D_{\bar n}}
\,\, : \,\, \Pi_{\bar n}(\underline x) = \o_{0,\g}\big|_{D_{\bar n}}
   $$
We have to characterize the family $\FF_{\bar n}$ of the (weak) limit 
points of $\{\nu_\g = \Pi_{\bar n}(\bar f_\g d\mu_\g);\, \g\in(0,1]\}$. 
First of all we observe that, by translation invariance, for any
finite region $\L$ of $D_{\bar n}$, 
   $$
\E^{\nu_\g}\big[N_\L\big] = N\g^d |\L| \le |\L|
   \Eq(p45)
   $$
This proves that the family $\{\nu_\g = \Pi_{\bar n}(\bar f_\g d\mu_\g) 
;\, \g\in(0,1]\}$ is tight.

Let $\mu_{m,\bar n}^{\bar \o}$ be the canonical Gibbs measure on the
cube $D_{\bar n}$ with boundary conditions $\bar \o\in \O$ and number 
of particles $m$. We prove that any $\nu \in \FF_{\bar n}$ can 
be written as 
   $$
\nu(d\o) = \int\!\hat\nu (d\bar\o, dm)\, \mu_{m,\bar n}^{\bar \o}(d\o) 
   \Eq(p46) 
   $$
where $\hat\nu (d\bar\o, dm)$ is a measure supported on $\{m\le
|D_{\bar n}|\}$. Since any limit point satisfies an inequality like 
\equ(p45), we have only to prove \equ(p46) for some $\hat \nu$.
By a straightforward extension to higher dimensions of the analogous 
argument in  Varadhan, [\rcite{V}, Lemma 7.5], we can reduce the problem 
to the estimate of a certain Dirichlet form. More precisely \equ(p46)
follows if
   $$
\sup_{\g\in (0,1]} \g^d \int\!d\mu_\g \, \sum_i {|\nabla_i \bar f_\g|^2 
\over \bar f_\g} < + \infty
   \Eq(p47)
   $$
The bound \equ(p47) is consequence of the following lemma:

\goodbreak
\vskip.2truecm
\noindent{\bf \Lemma (sp7)} {\sl Let
   $$
\bar f_\g(t,\underline x) \doteq {1\over t}\int_0^t\!ds\, 
\int_{\TT^d}\!dx\, f_\g(t,\t_x\underline x)
   \Eq(p48bis)
   $$
and define, for any density $f$,
   $$
\s_\g(f) \doteq \g^d \int\!d\mu_\g \, \sum_i 
{|\nabla_i f|^2 \over f}
   $$
Then there is $C>0$ such that, for any $\g\in (0,1]$ and 
any $t\in (0,T]$,
   $$
\s_\g(\bar f_\g(t)) \le {C\over t}
   \Eq(p49)
   $$
}
\vskip.2truecm

\noindent {\it Proof.} With an abuse of notation, denote by 
$H(f_\g(t)|1)$ the relative entropy of $f_\g(t)d\mu_\g$ 
w.r.t. $d\mu_\g$. Observing that $L_\g 1 = 0$ we get, after some 
standard computations,
   $$
\eqalign{
{d\over dt}H(f_\g(t)|1) & = {d\over dt} \int\!d\mu_\g\, f_\g(t)
\log f_\g(t) = \int\!d\mu_\g\, L_\g^* f_\g(t) \log f_\g(t) 
\cr & = - \g^{-d} \s_\g (f_\g(t)) - \g^d \int\!d\mu_\g\, 
\sum_{i\ne j} \b\nabla J(x_i-x_j) \cdot \nabla_i f_\g (t)
}
   \Eq(p50)
   $$
Since for any $x\in \TT^d$ $\t_x f_\g (t) = f_\g(t,\t_x\cdot)$ solves
the same Fokker-Planck equation, recalling definition \equ(p43), 
from \equ(p50) we get 
   $$
\eqalign{
\int_{\TT^d}\! dx\, {H(\t_x f_\g(t)|1) - H(\t_x f_\g(0)|1) \over t}  
= & - {1\over t} \int_0^t\! ds \int_{\TT^d}\! dx\, \g^{-d} \s_\g 
(\t_x f_\g(s))\cr & - \int\!d\mu_\g\, \g^d \sum_{i\ne j} 
\b\nabla J(x_i-x_j) \cdot \nabla_i \bar f_\g(t) 
}
   \Eq(p51)
   $$
Now, since $\s_\g(\cdot)$ is a convex functional, [\rcite{DS}], 
   $$
\s_\g(\bar f_\g(t)) \le {1\over t}\int_0^t\! ds 
\int_{\TT^d}\! dx\, \s_\g (\t_x f_\g(s))
   \Eq(p52)
   $$
On the other hand, by Cauchy-Schwartz inequality,
   $$ 
\eqalign{&
- \int\!d\mu_\g\, \g^d \sum_{i\ne j} \b\nabla J(x_i-x_j) 
\cdot \nabla_i \bar f_\g(t) \cr & \le 
\sqrt{\int\!d\mu_\g\, \bar f_\g(t) \g^d \sum_i \bigg|
\sum_{j:j\ne i} \b\nabla J(x_i-x_j) \bigg|^2} 
\sqrt{\s_\g(\bar f_\g(t))} 
\le  C_1 \g^{-d}\sqrt{\s_\g(\bar f_\g(t))}
}
   \Eq(p53)
   $$
with $C_1 = \b\|\nabla J\|_\infty$. Also, since $d\mu_\g$ 
is $\t_x$-invariant, 
   $$
\eqalign{
H(\t_x f_\g(0)|1) & = H(f_\g(0)|1) = \g^{-d} H_\g(0)
-\log C_\g(0) \cr & + \int\! d\mu_\g\, f_\g (0,\underline x) 
\sum_i \b\l(0,x_i) \le \g^{-d} H_\g(0) + 2 \b\|\l(0,\cdot)\|_\infty 
N \le C_2 \g^{-d}
}
   \Eq(p54)
   $$
for some $C_2>0$ (in the first bound we used Jensen's inequality, in
the second one the assumption \equ(p11) on the initial distribution). 
Collecting together \equ(p51), \equ(p52), \equ(p53) and \equ(p54), 
recalling also that the relative entropy is a positive function, we
get
   $$
-{C_2\over t}\g^{-d} \le - \g^{-d} \s_\g(\bar f_\g(t)) + C_1 
\g^{-d}\sqrt{\s_\g(\bar f_\g(t))}
   $$
so that, for any $\g\in (0,1]$,
   $$
\s_\g(\bar f_\g(t)) \le C_1 \sqrt{\s_\g(\bar f_\g(t))} +
{C_2 \over t} 
   \Eq(p55)
   $$
But \equ(p55) implies that $\s_\g(\bar f_\g(t)) \le C/t$ with $C$ 
the positive solution of $x=C_1\sqrt{Tx} + C_2$. The lemma is proven. 
\qed

\vskip.2truecm
Now we conclude the proof of Theorem \equ(sp6). Using
\equ(p46), the l.h.s. of \equ(p44) can be bounded by
   $$
\limsup_{n\to\infty}\sup_{\mu\in\GG_1} \E^\mu \Big[\big| 
F_n(\o) - \hat F(R_n(\o))\big| \1_K(R_n(\o)) \Big] 
   $$
where $\GG_1$ is the class of Gibbs states with density $\rho \le 1$.
The characteristic function $\1_K$ reduces the problem to computing the
above limit in the one phase region. The limit is then zero by the law
of large numbers for the unique Gibbs state of given density 
$\rho \in K$.

\vskip.2truecm

Finally, the limit \equ(p41) follows easily from Theorem \equ(sp6). In
fact \equ(p42) with $F=\phi_k\circ G^{\xi\eta}$ implies that the l.h.s. 
of \equ(p41) can be bounded by
   $$
\lim_{k\to \infty} T \sup_{\rho\in K} \big\{\big| \widehat 
{\phi_k \circ G^{\xi\eta}}(\rho) - \hat G^{\xi\eta}(\rho)\big|\big\}
   \Eq(p56)
   $$
But, for any $\rho \in K$,
   $$
\widehat {\phi_k \circ G^{\xi\eta}}(\rho) - \hat G^{\xi\eta}(\rho) =
\E^{\mu_\rho} \Big[(\phi_k \circ G^{\xi\eta})(\o) - G^{\xi\eta}(\o)
\Big]
   $$
where $\mu_\rho$ is the (unique) Gibbs state with chemical potential
$\l = \partial_\rho a (\b,\rho)$. We observe now that 
$\phi_k \circ G^{\xi\eta} \to G^{\xi\eta}$ pointwise as $k\to
\infty$. Moreover $|\phi_k \circ G^{\xi\eta}| \le |G^{\xi\eta}|\le 
c N_B^2$ for some $c>0$ and some finite subset $B$ of
$\R^d$. Recalling that, by superstability, 
$\E^{\mu_\rho}[N_B^2(\o)]<\infty$, the limit \equ(p56) is zero 
by the Dominated Convergence Theorem.

\vskip.2truecm

We conclude the section by proving Lemma \equ(sp4):

\vskip.2truecm

\noindent{\it Proof of Lemma \equ(sp4).} Since $V$ is positive and
superstable, there is a constant $\tilde C_W$ such that,
for any $\underline x \in \TT^{dN}$,
   $$
\sum_{i\ne j}W(\g^{-1}(x_i-x_j)) \le \tilde C_W
\sum_{i\ne j}V(\g^{-1}(x_i-x_j))
   $$
The previous inequality is a straightforward extension to higher 
dimensions of the analogous one derived in the proof of Lemma 4.2 
of [\rcite{V}], so we omit the details. Then it is enough to prove 
the lemma for $W=V$. From the basic entropy inequality, recalling 
definition \equ(p5) of the reference measure $\mu_\g$, 
   $$
\eqalign{
\E^{f_\g(t)}\bigg[\g^d\sum_{i\ne j}V(\g^{-1}(x_i-x_j))\bigg] & \le
- {2\g^d\over\b}\log\int\!d\underline x \, \exp\bigg[ - {\b\over 2}
\sum_{i\ne j}V(\g^{-1}(x_i-x_j))\bigg] \cr &
+ {2\g^d\over\b} H(f_\g(t)|1)
}
   $$
and, by Jensen inequality,
   $$
- {2\g^d\over\b}\log\int\!d\underline x \, \exp\bigg[-{\b\over 2}
\sum_{i\ne j}V(\g^{-1}(x_i-x_j))\bigg] \le
\g^d \int\! d\underline x \, \sum_{i\ne j}V(\g^{-1}(x_i-x_j))
\le \|V\|_\infty
   $$
Then we are left with an estimate of the relative entropy 
$H(f_\g(t)|1)$. Recalling \equ(p50) and that $\s_\g(\cdot)$ is a 
positive functional, we can bound
   $$
H(f_\g(t)|1) \le H(f_\g(0)|1) - \g^d \int_0^t\! ds 
\int\!d\mu_\g\, \sum_{i\ne j} \b\nabla J(x_i-x_j) 
\cdot \nabla_i f_\g (s)
   \Eq(p81)
   $$
Since $\mu_\g (d\underline x)$ is translation invariant,
recalling \equ(p48bis) and using \equ(p53), we have
   $$
- \g^d \int_0^t\! ds \int\!d\mu_\g\, \sum_{i\ne j} 
\b\nabla J(x_i-x_j) \cdot \nabla_i f_\g (s) \le 
C_1 t \g^{-d} \sqrt{\s_\g(\bar f_\g(t))}
   \Eq(p82)
   $$
The r.h.s. of \equ(p82) can be bounded using \equ(p49).
Then, recalling \equ(p54), from \equ(p81) and \equ(p82)
we finally get, for some $\tilde C >0$,
   $$
{2\g^d\over\b} H(f_\g(t)|1) \le \tilde C 
   \Eq(p83)
   $$
The lemma is proved. \qed

\goodbreak
\vskip1truecm
\centerline {\bf 5. Large deviation estimates and removal 
of the cutoffs.}

\vskip.5truecm
\numsec= 5
\numfor= 1
\numtheo=1

Part of the large deviation estimates of this section are contained in
the theory developed in [\rcite{GZ1}], [\rcite{GZ2}] and [\rcite{OVY}]. 
We will sometimes refer to these papers for proofs and details.

Let us start with some elementary facts in the theory of point 
processes. A configuration of particles in $\R^d$ can be represented
by a locally finite subset $\o$ of $\R^d$. Sometimes it can be useful 
to look at $\o$ as a Radon point measure on $\R^d$ via the map
$\o \mapsto \sum_{q\in\o}\d(\cdot - q)$. We denote by $\O$ the set of
all such configurations. $\O$ can be made into a Polish space under
the vague topology $\t_\O$, defined as the smallest topology making
continuous the mappings $\o \mapsto \int\!\o(dq)g(q) = 
\sum_{q\in\o}g(q)$ for any $g:\R^d\to\R$ which is continuous and 
compactly supported.
The natural $\s$-algebra $\FF$ on $\O$ is the one generated by
the counting variables $N_B:\o\to{\rm card}(\o\cap B)$ for $B$ any
Borel subset of $\R^d$. It can be proven that $\FF$ is the Borel
$\s$-algebra relative to the topology $\t_\O$. 

A point process on $\R^d$ is a probability measure $Q$ on
$(\O,\FF)$. We denote by $\MM$ the set of all point processes 
$Q$ with finite expected number of particles $\E^Q[N_B]$ in any
bounded Borel set $B\subset\R^d$. $\MM$ can be equipped with the 
topology $\t_w$ of weak convergence based on the topology $\t_\O$.
However it is useful to introduce a finer topology on $\MM$, called
the topology $\t_{\LL}$ of local convergence. Let $\LL$ be the 
class of measurable functions $F$ on $\O$ that are local and tame,
i.e. for any such $F$ there are a bounded set $B$ and a constant
$c>0$ such that $F(\o) = F(\o\cap B)$ and $|F(\o)|\le c(1+N_B(\o))$.
Then $\t_{\LL}$ is defined as the weak$^*$ topology on $\MM$ relative 
to $\LL$, i.e. the smallest one making continuous the mappings
$Q\mapsto Q(F) \doteq \E^Q[F]$ for any $F\in \LL$. Note that
$\t_{\LL}$ is strictly finer than $\t_w$, as follows observing that 
the mappings $Q \mapsto Q(N_B)$ are $\t_\LL$-continuous for any
bounded Borel set $B$.

Let $\MM_\t$ be the set of all the stationary point processes $Q$ in
$\MM$, i.e. those $Q$ such that $Q=Q\circ\t_x^{-1}$ for any 
$x\in\R^d$. $\MM_\t$ is $\t_{\LL}$-closed and is assumed equipped with 
the induced topology.
For any $Q\in\MM$ there is a Radon measure $\nu_Q(dx)$ on $\R^d$
such that $Q(N_B)=\nu_Q(B)$ for any Borel set $B\subset\R^d$.
If $Q\in\MM_\t$ then $\nu_Q(dx) = \rho(Q) dx$ for some positive number
$\rho(Q)$, called the intensity of $Q$.

Let $P$ be the Poisson point process on $\R^d$ with intensity 
$\rho(P)=1$ (i.e. $P$ is such that for any collection of disjoint 
bounded subsets $B_1,\ldots,B_n$, the counting variables 
$N_{B_1},\ldots,N_{B_n}$ are independent and Poisson distributed with 
parameters $|B_1|,\ldots,|B_n|$). 
We introduce the entropy density $h(Q|P)$ of $Q\in \MM_\t$ 
w.r.t. $P$ as follows. For any $B\subset \R^d$ we denote by $\pi_B$ the 
projection on $B$, and let $D_n$, $n \in \N$, be as in definition 
\equ(p23bis). Denote by $\HH_n(Q|P)$ the relative entropy of 
$Q\circ \pi_{D_n}^{-1}$ w.r.t. $P\circ\pi_{D_n}^{-1}$. It is easy
to see that $n \mapsto \HH_n(Q|P)$ is a super-additive functional, so we
can define
   $$
h(Q|P) = \lim_{n\to\infty}|D_n|^{-1}\HH_n(Q|P) = \sup_n|D_n|^{-1}
\HH_n(Q|P)
   \Eq(p57)
   $$

Let now $V$ be a positive and superstable finite range potential like
the one introduced in Section 2. The associated Hamiltonian in $D_n$ 
with free boundary conditions is 
   $$
H_n(\o) = {1\over 2} \sum_{\scriptstyle q,q' \in \o \cap D_n 
\atop \scriptstyle q \ne q'} V(q-q'), \quad \o\in\O
   \Eq(p58)
   $$
For each $n\in \N$ and $Q\in\MM_\t$ let
   $$
\Phi_n(Q) \doteq |D_n|^{-1} Q(H_n)
   $$
be the expected energy per volume in $D_n$. By our assumptions on 
the potential, $\Phi_n$ is well defined and positive. Moreover, see
[\rcite{GZ2}, Thm. 1], the limit $\Phi(Q) = \lim_{n\to\infty}
\Phi_n(Q)$ exists and satisfies
   $$
\Phi (Q) = \cases{ Q(U) & if  $Q\in\MM_{\t,2} $ \cr 
                   +\infty & otherwise \cr}
   \Eq(p59)
   $$
where 
   $$
\MM_{\t,2} \doteq \{Q\in\MM_\t\,:\, Q(N_B^2)< + \infty 
\text{for any bounded} B\subset\R^d\}
   $$
and 
   $$
U(\o) = {1\over 2} \sum_{\scriptstyle q,q' \in \o 
\atop \scriptstyle q \ne q'} \chi (q) V(q-q') 
   \Eq(p60)
   $$
In \equ(p60) $\chi$ is any non negative function on $\R^d$ with
compact support and total integral 1 (it is easy to verify that
$Q(U)$ does not depend on $\chi$ if $Q$ is a stationary point
process).

Finally define, for any $\l\in\R$ and $\b\ge 0$,  
   $$
K_{\b,\l}(Q) = \b \Phi(Q) + h(Q|P) - \b\l\rho (Q)
   \Eq(p61)
   $$

The following proposition establishes the basic properties of the 
functionals introduced above (see [\rcite{GZ1}], [\rcite{GZ2}] and
[\rcite{OVY}]). 

\goodbreak
\vskip.2truecm
\noindent{\bf \Proposition (sp8)} {\sl The functionals $\Phi, 
K_{\b,\l} : \MM_\t \to [0,+\infty]$ are lower semicontinuous relative 
to $\t_{\LL}$ (and then also relative to any coarser topology on
$\MM_\t$, e.g. $\t_w$). Moreover $K_{\b,\l}$ has $\t_{\LL}$-compact 
level sets.
}
\vskip.2truecm

We can formulate now the LDP for Gibbsian point processes. The form 
we need here is strictly contained in the results of 
the papers quoted above:

\goodbreak
\vskip.2truecm
\noindent{\bf \Theorem (sp9)} {\sl Let $F\in \LL$ and $F_n$ be as in 
definition \equ(p23bis). Then, for any $\l\in\R$ and $\b\ge 0$, there 
exists the limit
   $$
\lim_{n\to\infty}|D_n|^{-1} \log \E^P \exp\big[\b\l N_{D_n} - 
\b H_n + |D_n| F_n\big] = \sup_{Q\in\MM_\t}\{Q(F) - K_{\b,\l}(Q)\}
   \Eq(p62)
   $$
and the r.h.s. of \equ(p62) is finite.
}
\vskip.2truecm

\noindent {\it Remark.} In [\rcite{GZ2}] the principle is formulated 
in terms of functionals of the stationary empirical field $R_{n,\o}$ 
obtained by replacing the configuration $\o$ with the periodic 
continuation $\o^{(n)}$ of its restriction to $D_n$, but the same 
result holds also for the spatial average $F_n$ by standard arguments
that we do not describe here, see e.g. [\rcite{GZ1}]. 

\vskip.2truecm

An immediate consequence of the above principle is the existence of
the pressure $p(\b,\l)$ and a variational formula for it. In fact,
taking $F\equiv 0$ in \equ(p62) and recalling \equ(p7) and
\equ(p7bis), we get
   $$
p(\b,\l) = - \min_{Q\in\MM_\t} \b^{-1} K_{\b,\l}(Q)
   \Eq(p63)
   $$
and, comparing with \equ(p8), we recover the Helmholtz free energy as
   $$
a(\b,\rho) = \min_{Q\in\MM_\t:\rho(Q)=\rho} \{\Phi(Q) +
\b^{-1} h(Q|P)\}
   \Eq(p64)
   $$
where we used the fact that the r.h.s. of \equ(p64) is a convex
function of $\rho$ (which is true since $\Phi$ and $h(\cdot|P)$ are 
affine functionals on $\MM_\t$).

Now we discuss the LDP for local Gibbs states of the type defined in 
\equ(p10). We forget the dependence on $t\in [0,T]$ and we deal 
with a generic smooth map $x\mapsto\l(x)$ of $\TT^d$  
into the one phase region $\UU$ such that the corresponding 
density $x\mapsto\rho (x)\in\WW$ satisfies
   $$
\int_{\TT^d}\!dx\, \rho (x) = 1
   \Eq(p65)
   $$
We denote by $\hat f_\g (\underline x)$ the density w.r.t. 
$\mu_\g(d\underline x)$ of the associated local Gibbs state
(observe that the functions $x\mapsto\l(t,x)$ introduced in Section 2 
satisfy the conditions above for any $t\in [0,T]$). 
The following theorem is contained in Section 5 of [\rcite{OVY}].

\goodbreak
\vskip.2truecm
\noindent{\bf \Theorem (sp10)} {\sl Let $A \in \LL$ and 
$\vf\in C(\TT^d)$. Then
   $$
\eqalign{
\lim_{\g\downarrow 0} & \g^d \log  \int\!d\underline x\, 
\exp \bigg[ \sum_i \b\l(x_i) - {\b\over 2}\sum_{i\ne j} 
V(\g^{-1}(x_i-x_j)) +\g^{-d} \int_{\TT^d} \!dx\, \vf(x) 
A(\o_{x,\g}) \bigg] \cr &
= \sup\bigg\{\int_{\TT^d} \!dx\, \big[\vf(x) Q_x(A) - K_{\b,\l(x)}
(Q_x)\big];\,\{Q_x\}\subset \MM_\t :\int_{\TT^d}\!dx\,\rho(Q_x) 
= 1 \bigg\}
}
   \Eq(p66)
   $$
}
\vskip.2truecm

Taking $A \equiv 0$ in \equ(p66) and recalling \equ(p63), we get
   $$
\eqalign{&
\lim_{\g\downarrow 0} \g^d  \log \int\!d\underline x\, 
\exp \bigg[ \sum_i \b\l(x_i) - {\b\over 2}
\sum_{i\ne j} V(\g^{-1}(x_i-x_j)) \bigg] \cr &
= \sup\bigg\{\int_{\TT^d} \!dx\, \big[-K_{\b,\l(x)}(Q_x)\big];
\,\{Q_x\}: \int_{\TT^d}\!dx\,\rho(Q_x) = 1 \bigg\} 
= \int_{\TT^d}\!dx\, \b p(\b,\l(x))
}
   \Eq(p67)
   $$
where in the last equality we used the fact that the supremum of 
the integrand is reached at $\rho(Q_x)=\rho(x)$ 
satisfying \equ(p65).

Theorem \equ(sp10) is not sufficient for our purposes since 
to prove \equ(p38) we need also an upper bound for the Laplace 
asymptotics of non local functionals. The following theorem
gives the required estimate.

\goodbreak
\vskip.2truecm
\noindent{\bf \Theorem (sp11)} {\sl Let $A,F,G \in \LL$ with $F$ 
bounded and let $\vf\in C(\TT^d)$ and $\psi\in C(\TT^d\times\TT^d)$. 
Then:
   $$
\eqalign{
\limsup _{\g\downarrow 0} \g^d & \log \E^{\hat f_\g}
\exp \g^{-d}  \bigg[\int_{\TT^d} \!dx\, \vf(x) A(\o_{x,\g}) 
+ \int_{\TT^d} \!dx \int_{\TT^d} \!dy\, \psi(x,y)
F(\o_{x,\g})G(\o_{y,\g}) \bigg] \cr &
\le \sup\bigg\{\int_{\TT^d}\!dx\,\int_{\TT^d}\!dy\, 
\big[\vf(x) Q_x(A) + \psi(x,y) Q_x(F)Q_y(G) \cr &
- K_{\b,\l(x)}(Q_x) - \b p(\b,\l(x))\big];\,\{Q_x\}\subset\MM_\t
: \int_{\TT^d}\!dx\, \rho(Q_x) = 1 \bigg\}
}
   \Eq(p68)
   $$
}
\vskip.2truecm

\noindent {\it Proof.} For any configuration of particles 
$\o$ on $\TT^d$ define 
   $$
\eqalign{
T(\o) & = \sum_{z\in\o}\b\l(z) - {\b\over 2}
\sum_{\scriptstyle z,z' \in \o \atop \scriptstyle z \ne z'} 
V(\g^{-1}(z-z')) +\g^{-d}\int_{\TT^d}\!dx\,\vf(x) A(\o_{x,\g}) 
\cr & + \g^{-d} \int_{\TT^d} \!dx \int_{\TT^d} \!dy\, \psi(x,y)
F(\o_{x,\g})G(\o_{y,\g})
}
   \Eq(p69)
   $$
Setting
   $$
Z_N^c \doteq \int\!d\underline x\, \exp[T(\underline x)]
   $$    
and making use of \equ(p67) it is enough to prove that
   $$
\eqalign{
\limsup_{\g\downarrow 0} \g^d \log Z_N^c
\le \sup\bigg\{& \int_{\TT^d}\!dx\,\int_{\TT^d} 
\!dy\,  \big[\vf(x) Q_x(A) +\psi(x,y) Q_x(F) Q_y(G) \cr &
- K_{\b,\l(x)}(Q_x)\big];\,\{Q_x\}\subset\MM_\t
: \int_{\TT^d}\!dx\, \rho(Q_x) = 1 \bigg\}
}
   \Eq(p70)
   $$

First of all we observe that for any partition of $\TT^d$ into
cubes of side $2\e$ there are uniform approximations 
of $\l,\vf$, and $\psi$ that are constant on this partition. Let
$T_\e(\cdot)$, $Z_{N,\e}^c$ be defined as $T(\cdot)$, $Z_N^c$
with $\l,\vf$, and $\psi$ replaced by their approximations. Let
$D_{\ell_0}$ be a cube (as in definition \equ(p23bis)) such that 
the functions $A$, $F$ and $G$ depend only on $\o\cap D_{\ell_0}$ 
and, for some $c>0$, $A,G \le c(1+N_{D_{\ell_0}})$. Then one easily
bounds
   $$
|T(\underline x) - T_\e(\underline x)| \le
O_1(\e) \g^{-d} \int_{\TT^d}\!dx\, \big(1+N_{D_{\ell_0}}
(\o_{x,\g})\big) \le O_1(\e)|D_{\ell_0}|(1+N)
   \Eq(p701)
   $$
with $O_1(\e)\to 0$ as $\e\downarrow 0$ (uniformly in $\g$), so that
   $$
\limsup_{\g\downarrow 0}\g^d\log Z_N^c \le \limsup_{\g\downarrow 0}
\g^d\log Z_{N,\e}^c + O_1(\e)|D_{\ell_0}|
   $$
On the other hand, also the error made in replacing the r.h.s. of
\equ(p70) with its $\e$-approximation (i.e. the one defined with
the piecewise constant approximations of $\l,\vf$, and $\psi$) is
easily bounded by $O_2(\e)|D_{\ell_0}|$ with $O_2(\e)\to 0$ as 
$\e\downarrow 0$. Thus, with no loss of generality we can prove
\equ(p70) assuming $\l,\vf$, and $\psi$ constant on some cubic 
partition of $\TT^d$.

Let $P_\g$ be the Poisson point process on $\TT^d$ with intensity 
$\g^{-d}$ and define, for any $\mu\in\R$,
   $$
Z_\mu(\g) \doteq \E^{P_\g}\exp[T(\o)+\mu N(\o)] = 
e^{-\g^{-d}} \sum_{n=0}^\infty 
{\g^{-dn}e^{\mu n} \over n!} Z_n^c
   $$
Since 
   $$
\lim_{N\to\infty} {1\over N} \log{e^{-N}N^N \over N!} = 0
   $$
then (recall that $N\g^d\nearrow 1$ as $\g\downarrow 0$)
   $$
\limsup_{\g\downarrow 0}\g^d \log Z_N^c \le \inf_{\mu\in\R}
\limsup_{\g\downarrow 0}\big\{\g^d\log Z_\mu (\g) - \mu\big\}
   \Eq(p70bis)
   $$
From \equ(p70bis) we get \equ(p70) if
   $$
\eqalign{
\limsup_{\g\downarrow 0} \g^d \log Z_\mu(\g) 
\le\sup_{\{Q_x\}\subset\MM_\t}\int_{\TT^d}\!dx\,\int_{\TT^d} 
\!dy\, & \big\{\vf(x) Q_x(A) + \psi(x,y) Q_x(F) Q_y(G) \cr & 
- K_{\b,\l(x)}(Q_x) + \mu\rho (Q_x) \big\}
}
   \Eq(p71)
   $$

By redefining $\l(x)$ as $\l(x) + \mu$, it is enough to prove 
\equ(p71) for $\mu=0$. Divide $\TT^d$ into disjoint boxes $B_\s$ of 
center $\s$ and side $(2\g\ell)$ and let $D_{\ell_0}$ be the cube
as in \equ(p701). Set
   $$
B_{\s,\ell_0} = \{ x\in B_\s : {\rm dist}(x,\TT^d\setminus B_\s) 
\ge 2\g\ell_0\}
   \Eq(p71bis)
   $$
and define
   $$
\eqalign{
T_{\s\s'}(\o) & = (2\g\ell)^d \bigg[\b\l (\s) N_{B_\s} - 
{\b\over 2} \sum_{\scriptstyle z,z' \in \o \cap B_\s \atop
\scriptstyle z \ne z'} V(\g^{-1}(z-z')) +\g^{-d}\vf(\s) 
\int_{B_{\s,\ell_0}} \!dx\, A(\o_{x,\g}) \bigg] \cr &
+ \g^{-d}\psi(\s,\s') \int_{B_{\s,\ell_0}} \!dx 
\int_{B_{\s',\ell_0}} \!dy\, F(\o_{x,\g})G(\o_{y,\g})
}
   \Eq(p72)
   $$

Since the potential is positive, we obtain an upper bound on
$T(\o)$ by neglecting the interaction between different boxes.
Then 
   $$
T(\o) \le \sum_{\s,\s'} T_{\s\s'}(\o) + {\cal R} (\o)
   \Eq(p73)
   $$
where ${\cal R}$ takes into account the errors due to the integration
on the smaller cubes $B_{\s,\ell_0}$ in \equ(p72). Because of the choice
of $\ell_0$ we easily bound, for some $C>0$,
   $$
{\cal R} \le C\g^{-d} \int\limits_{\cup (B_\s\setminus B_{\s,\ell_0})}
\!dx\, \big(1 + N_{D_{\ell_0}}(\o_{x,\g})\big)
   $$

We decompose $\bigcup (B_\s\setminus B_{\s,\ell_0})$ into a union of 
disjoint cubes of side $(2\g\ell_0)$. Calling $\hat x_\a$ the 
centers of these cubes, we have
   $$
{\cal R} \le C \sum_{\a=1}^n \int_{D_{\ell_0}} \!dr\, 
\big(1 + N_{D_{\ell_0}}(\t_{r+\g^{-1}\hat x_\a}\o_{0,\g})\big)
   $$
with $n \sim (\g\ell)^{-d}(\ell/\ell_0)^{d-1}$ so that, by Jensen
inequality,
   $$
\limsup_{\ell\to\infty}\limsup_{\g\downarrow 0} 
\g^d\log\E^{P_\g} \exp [{\cal R}] \le
\limsup_{\ell\to\infty}\limsup_{\g\downarrow 0}\g^d n\log\E^P
\exp[C(1+N_{D_{\ell_0}})] = 0
   \Eq(p74)
   $$
where we used the independence and stationarity of $P_\g$ and
that $\E^P\exp[aN_B]<\infty$ for any $a>0$ and any bounded set 
$B\subset\R^d$. From \equ(p73) and \equ(p74) we get
   $$
\limsup_{\g\downarrow 0} \g^d \log Z_0(\g) \le
\limsup_{\ell\to\infty}\limsup_{\g\downarrow 0}
\g^d \log\E^{P_\g}\exp\bigg[\sum_{\s,\s'}T_{\s\s'}(\o)\bigg]
   \Eq(p75)
   $$

Now define the following local function on $\O\times\O$: 
   $$
S_{\s\s'} (\o_1,\o_2) = \b\l(\s) R(\o_1) - \b U(\o_1) + 
\vf(\s) A(\o_1) + \psi(\s,\s')F(\o_1)G(\o_2) 
   $$
where $R$ and $U$ are defined in \equ(p24) and \equ(p60) respectively. 
Expanding variables in the r.h.s. of \equ(p75) and neglecting errors 
that can be proved to be of order $O(1/\ell)$ uniformly in
$\g\in(0,1]$ as done in \equ(p74), one easily obtains (introducing a
new parameter $\ve =2\ell\g$)
   $$
\limsup_{\g\downarrow 0} \g^d\log Z_0(\g) \le \limsup_{\ell\to\infty}
\limsup_{\ve\downarrow 0} \XX(\ell,\ve)
   \Eq(p77)
   $$
with
   $$
\XX(\ell,\ve) \doteq
\ve^d |D_\ell|^{-1} \log\E^{
\otimes_{\s} P_\s}\exp\bigg[\ve^d \sum_{\s,\s'} |D_\ell| 
\big(S_{\s\s'}\big)_\ell (\o_\s,\o_{\s'})\bigg]
   \Eq(p771)
   $$
where $\otimes_\s P_\s$ is the product measure on 
$\O^{(\ve^{-d})}= \O\times\cdots\times \O$ of $\ve^{-d}$ 
independent Poisson processes $P_\s$ on $\R^d$ and,
analogously to \equ(p23bis), for any local function $M$ on
$\O\times\O$, we defined
   $$
M_\ell(\o_1,\o_2) \doteq
{1\over |D_\ell|^2}\int_{D_\ell}\!dr\int_{D_\ell}\!dr'\,
M(\t_r\o_1,\t_{r'}\o_2)
   $$  
To analyze the limit in the r.h.s. of \equ(p77) we need first to
introduce some cutoffs. Given $k_+,k_-\in\N$, let $\XX_{k_+k_-}
(\ell,\ve)$ be defined as $\XX(\ell,\ve)$ in \equ(p771) with 
$S_{\s\s'}$ replaced by
   $$
S_{\s\s'}^{k_+k_-} \doteq \cases{k_+ & if $S_{\s\s'}>k_+$ \cr 
                      S_{\s\s'} & if $-k_-\le S_{\s\s'}\le k_+$
                       \cr -k_- & if $S_{\s\s'}<-k_-$}
   $$ 
Then, by definition of relative entropy,
   $$
\XX_{k_+k_-}(\ell,\ve) \le \sup_Q \bigg\{ \E^Q\bigg[\ve^{2d}
\sum_{\s,\s'} \big(S_{\s\s'}^{k_+k_-}\big)_\ell(\o_\s,\o_{\s'})
\bigg] - \ve^d |D_\ell|^{-1} H\big(Q\big|\otimes_\s 
(P_\s \circ\pi_{D_{\ell+\bar\ell}}^{-1})\big)  \bigg\} 
   \Eq(p78)
   $$
where the supremum is taken over point processes $Q$ on 
$D_{\ell+\bar\ell}\times\cdots\times D_{\ell+\bar\ell}$.
The parameter $\bar\ell$ is chosen so large that $S_{\s\s'}$
is $D_{\bar\ell}$-measurable. The variational
formula for the relative entropy gives also the explicit form of 
the measure $\bar Q$ where the supremum in the r.h.s. of \equ(p78) 
is achieved, see e.g. [\rcite{DE}, Prop. 1.4.2]: $\bar Q
\ll \otimes_\s \big(P_\s\circ\pi_{D_{\ell+\bar\ell}}^{-1}\big)$ and
   $$
{d\bar Q\over d\big(\otimes_\s \big( 
P_\s\circ\pi_{D_{\ell+\bar\ell}}^{-1}\big)\big)} = {\cal N}_\ve 
\exp\bigg[\ve^d \sum_{\s,\s'}
|D_\ell|\big(S_{\s\s'}^{k_+k_-}\big)_\ell (\o_\s,\o_{\s'})
\bigg]
   \Eq(p79) 
   $$
(${\cal N}_\ve$ the normalization constant).
For any $x\in\TT^d$ let $\s(x)$ be such that $x\in B_{\s(x)}$, and 
denote by $Q_\s$, $Q_{\s\s'}$ the $\s$-th, $\{\s,\s'\}$-th marginals
of $Q$. From \equ(p79) it is easy to prove that, for any
$x,y\in\TT^d$, $x\ne y$,
   $$
\lim_{\ve\downarrow 0} D_{\rm var}\bigg(\bar Q_{\s(x)\s(y)}
(d\o_1,d\o_2), \int\bar Q(d\bar \o) \bar Q_{\s(x)}\big(d\o_1\big|
\bar\o_{\{\s(x)\}^c}\big) \bar Q_{\s(y)}\big(d\o_2\big|
\bar\o_{\{\s(y)\}^c}\big)\bigg) = 0
   $$
where $D_{\rm var}(\cdot,\cdot)$ denotes the variation distance 
between measures. Moreover 
   $$
H\big(Q\big|\otimes_\s \big(P_\s \circ\pi_{D_{\ell+\bar\ell}}^{-1}
\big)\big) \ge \sum_\s H\big(Q_\s \big| P_\s \circ
\pi_{D_{\ell+\bar\ell}}^{-1} \big)
   $$ 
Since $\l$, $\vf$, $\psi$ are piecewise constant, as $\ve$ is small
enough, for any $x,y\in\TT^d$, the function $S_{\s(x)\s(y)}$ is
independent on the partition $\{B_\s\}$ and equal to
   $$
S_{xy}(\o_1,\o_2) \doteq \b\l(x) R(\o_1) - \b U(\o_1) + 
\vf(x) A(\o_1) + \psi(x,y)F(\o_1)G(\o_2) 
   $$
Thus the limit as $\ve\downarrow 0$ of the r.h.s. of \equ(p78) is 
easily bounded and we get
   $$
\limsup_{\ve\downarrow 0} \XX_{k_+k_-}(\ell,\ve) \le
\sup_{\{Q_x'\}} \int_{\TT^d}\!dx\int_{\TT^d}\!dy\,\Big\{ 
Q_x'\otimes Q_y'\big(S_{xy}^{k_+k_-}\big) - |D_\ell|^{-1}
H\big(Q_x'\big|P\circ\pi_{D_{\ell+\bar\ell}}^{-1}\big)\Big\}
   \Eq(p80)
   $$
where the supremum is taken over all the collections 
$\{Q_x';x\in \TT^d\}$ of point processes on $D_{\ell+\bar\ell}$.
Now the proof follows in a standard way, see e.g. [\rcite{GZ1}],
[\rcite{OVY}]. We extend $Q_x'$ to a point process $Q_x''$ on 
$\R^d$ by taking independent copies on all disjoint cubes translated 
of $D_{\ell+\bar\ell}$. Then we obtain from it a stationary process 
$Q_x^{(\ell)}$ by setting
   $$
Q_x^{(\ell)} = |D_{\ell+\bar\ell}|^{-1}\int_{D_{\ell+\bar\ell}}
\!dr\, \t_r Q_x''
   $$
From convexity of the relative entropy and the independence properties
of $P$ one easily proves that $h(Q_x^{(\ell)}|P) \le |D_{\ell+
\bar\ell}|^{-1}H\big(Q_x'\big|P\circ\pi_{D_{\ell+\bar\ell}}^{-1}\big)$.
Moreover, for any bounded local function $M$ on $\O\times\O$,   
$|Q_x^{(\ell)}\otimes Q_y^{(\ell)}(M) - Q_x'\otimes Q_y'(M_\ell)|\to
0$ as $\ell\to\infty$. Then from \equ(p80) we finally obtain 
   $$
\limsup_{\ell\to\infty}\limsup_{\ve\downarrow 0}
\XX_{k_+k_-}(\ell,\ve) \le \WW\big(\{S_{xy}^{k_+k_-}\}\big)
   \Eq(p801)
   $$
where, for any collection $\{M_{xy}\}$ of local functions on 
$\O\times\O$, we defined
   $$
 \WW\big(\{M_{xy}\}\big) \doteq \sup_{\{Q_x\}\subset\MM_\t} 
\int_{\TT^d}\!dx\int_{\TT^d}\!dy\,\Big\{ 
Q_x\otimes Q_y\big(M_{xy}\big) - h(Q_x|P)\Big\} 
   $$

To remove the cutoffs we can argue as in the proof of 
Theorem 5.2 in [\rcite{OVY}], then we just sketch the argument. Let 
$S_{\s\s'}^{k_+} = \min\{S_{\s\s'};k_+\}$ and define $S_{xy}^{k_+}$ 
and $\XX_{k_+}(\ell,\ve)$ accordingly. 
Using the boundness of $S_{\s\s'}^{k_+k_-}$ and that 
$h(\cdot|P)$ has compact level sets, one can prove that 
$\lim\limits_{k_-\to\infty}\WW\big(\{S_{xy}^{k_+k_-}\}\big)
\le \WW\big(\{S_{xy}^{k_+}\}\big)$. Then,
since $\XX_{k_+}(\ell,\ve) \le \XX_{k_+k_-}(\ell,\ve)$, from
\equ(p801) we get
   $$
\limsup_{\ell\to\infty}\limsup_{\ve\downarrow 0}\XX_{k_+}(\ell,\ve)
\le \WW\big(\{S_{xy}^{k_+}\}\big)
   \Eq(p802)
   $$
Now we are left with the upper cutoff. Recalling definition \equ(p771),
by Holder inequality, for any $p,q>1$ such that $p^{-1}+q^{-1}=1$, and 
any $k_+\in\N$,
   $$
\eqalign{
\XX(\ell,\ve) & \le {\ve^d |D_\ell|^{-1}\over p} \log\E^{
\otimes_{\s} P_\s}\exp\bigg[\ve^d \sum_{\s,\s'} |D_\ell| 
\big(pS_{\s\s'}^{k_+}\big)_\ell (\o_\s,\o_{\s'})\bigg] \cr &
+ {\ve^d |D_\ell|^{-1}\over q} \log\E^{\otimes_{\s} P_\s}
\exp\bigg[q\ve^d \sum_{\s,\s'} |D_\ell| 
\big(\big(S_{\s\s'}\big)_\ell - \big(S_{\s\s'}^{k_+}\big)_\ell\big)
(\o_\s,\o_{\s'})\bigg]
}
\Eq(p803)
   $$
To bound the first term in the r.h.s. of \equ(p803) we can use
\equ(p802) with $S_{\s\s'}^{k_+}$ replaced by $pS_{\s\s'}^{k_+}$.
To bound the second term we recall first that there are $C>0$ and 
a bounded subset $B$ of $\R^d$ such that $S_{\s\s'} (\o_1,\o_2) \le 
C \big(1+N_B(\o_1) + N_B(\o_2)\big)$ for any $\s,\s'$, so that
   $$
\eqalign{
\big(S_{\s\s'}-S_{\s\s'}^{k_+}\big)(\o_1,\o_2) & \le
\big[C \big(1+N_B(\o_1) + N_B(\o_2)\big)-k_+\big]_+ \cr &
\le {1\over 2}\big[C+2CN_B(\o_1)-k_+\big]_+ + 
{1\over 2}\big[C+2CN_B(\o_2)-k_+\big]_+ 
}
   $$
Then, setting
   $$
Y_{\ell,k_+}(\o) \doteq \exp\bigg[{q\over 2}\int_{D_\ell}\!dr\,
\big[C+2CN_B(\t_r\o)-k_+\big]_+\bigg]
   $$
we have
   $$
\eqalign{ 
\E^{\otimes_{\s} P_\s} & \exp\bigg[q\ve^d \sum_{\s,\s'} |D_\ell| 
\big(\big(S_{\s\s'}\big)_\ell - \big(S_{\s\s'}^{k_+}\big)_\ell\big)
(\o_\s,\o_{\s'})\bigg] \cr & 
\le \E^{\otimes_{\s} P_\s}\bigg[\prod_{\s,\s'}Y_{\ell,k_+}
(\o_\s)^{\ve^d}Y_{\ell,k_+}(\o_{\s'})^{\ve^d}\bigg] \cr &
= \E^P\Big[Y_{\ell,k_+}(\o)^{\ve^d}\Big]^{\ve^{-2d}}
\le \E^P\Big[Y_{\ell,k_+}(\o)\Big]^{\ve^{-d}}
}
   $$
so that, from \equ(p803),
   $$
\limsup_{\ell\to\infty} \limsup_{\ve\downarrow 0}\XX(\ell,\ve) 
 \le \lim_{k_+\to\infty} \WW\big(\{pS_{xy}^{k_+}\}\big) 
+ \lim_{k_+\to\infty} \lim_{\ell\to\infty} {1\over q|D_\ell|}
\log\E^P\Big[Y_{\ell,k_+}(\o)\Big]
   \Eq(p804)
   $$
The second term in the r.h.s. of \equ(p804) is zero,
see [\rcite{OVY}, Thm. 5.2]. Then, since the first term in 
\equ(p804) is bounded by $\WW\big(\{pS_{xy}\}\big)$, in the 
limit $p\downarrow 1$ we finally get
   $$
\limsup_{\ell\to\infty}\limsup_{\ve\downarrow 0}\XX(\ell,\ve)
\le \WW\big(\{S_{xy}\}\big)
   \Eq(p805)
   $$
But (recall the above definition of $S_{xy}$) 
$\WW\big(\{S_{xy}\}\big)$ is equal to the r.h.s. of \equ(p71) 
with $\mu=0$, which then follows from \equ(p77) 
and \equ(p805). The theorem is proved. \qed

\vskip.2truecm

Now we can give the missing proofs of Section 3.

\vskip.2truecm

\noindent{\it Proof of \equ(p34).}
From the basic entropy inequality, for any $\d_0>0$,
   $$
\E^{f_\g(t)} \bigg[ \int_{\TT^d}\! dx\, \b\O_p(t,x)\bigg] -
\d_0^{-1} H_\g(t) \le \d_0^{-1}\g^d\log\E^{\hat f_\g(t)}
\exp\bigg[ \d_0\g^{-d}\int_{\TT^d}\!dx\, \b\O_p(t,x)\bigg]
   \Eq(p84)
   $$
As before we forget the dependence on $t\in [0,T]$ and we deal 
with a generic smooth map $x\mapsto\l(x)\in K_u$ such that the 
corresponding density $\rho(x) = \partial_\l p(\b,\l(x))\in K_w$ 
satisfies \equ(p65) (the compact sets $K_u$ and $K_w$ have been 
introduced just after Theorem \equ(sp1)). Let $\hat f_\g$ be the 
density of the corresponding local Gibbs state. From \equ(p84) it 
is enough to prove that, for $\d_0$ small enough and uniformly in 
the choice of $x\mapsto\l(x)$ above,
   $$
\limsup_{k\to\infty}\limsup_{n\to\infty}\limsup_{\g\downarrow 0}
\d_0^{-1}\g^d\log\E^{\hat f_\g}\exp\bigg[\d_0\g^{-d}\int_{\TT^d}
\!dx\, \b\O_p(x)\bigg] = 0, \quad p=3,4
   \Eq(p84bis)
   $$
where $\O_p(x)$ is defined as in \equ(p33) but relative to the 
map $x\mapsto\l(x)$. We analyze the cases $p=3,4$ separately:
   
\vskip.2truecm

($p=3$). From the smoothness of $x\mapsto\l(x)$ and since 
$|\vf * R_n(\o_{\cdot,\g})|\le \|\vf\|_\infty$, for any $\vf
\in C(\TT^d)$ and any $\o_{\cdot,\g}$, there is $C_1>0$ such that
   $$
\O_3(x) \le C_1 \big( 1 + R_n(\o_{x,\g}) + 
(\phi_k\circ |G|)_n(\o_{x,\g})\big) (1-u_n(\o_{x,\g}))
   \Eq(p85)
   $$
where
   $$
|G|(\o) \doteq {\b\over 2} \sum_{\scriptstyle q,q' \in \o 
\atop \scriptstyle q \ne q'} \chi(q) |\nabla V|(q-q') |q-q'|
   \Eq(p85bis)
   $$
From Theorem \equ(sp10), \equ(p67) and \equ(p85) we get
the following bound:
   $$
\eqalign{&
\limsup_{k\to\infty}\limsup_{n\to\infty}\limsup_{\g\downarrow 0}
\d_0^{-1}\g^d \log\E^{\hat f_\g} \exp\bigg[
\d_0\g^{-d}\int_{\TT^d}\!dx\, \b\O_3(x)\bigg] \cr &
\le \limsup_{k\to\infty}\limsup_{n\to\infty} \sup\bigg\{
\int_{\TT^d}\!dx\, \Big[\d_0\b C_1 Q_x\Big(\big( 1 + R_n + 
(\phi_k\circ |G|)_n\big) (1-u_n)\Big) \cr &
- K_{\b,\l(x)}(Q_x) - \b p(\b,\l(x)) \Big];\, \{Q_x\}\subset 
\MM_\t: \int_{\TT^d}\!dx\, \rho(Q_x) = 1 \bigg\}
}
   \Eq(p86)
   $$
For any $Q_x\in\MM_\t$ let $\int\!\nu_x(de)\,Q_e$ be its ergodic 
decomposition. Recalling that $K_{\b,\l}(\cdot)$ is an affine 
functional, the r.h.s. of \equ(p86) becomes
   $$
\eqalign{ &
\limsup_{k\to\infty}\limsup_{n\to\infty}\sup\bigg\{\int_{\TT^d}
\!dx\,\int\!\nu_x(de)\, \Big[\d_0\b C_1 Q_e\Big(\big( 
1 + R_n + (\phi_k\circ |G|)_n \big)(1-u_n)\Big) \cr & 
- K_{\b,\l(x)}(Q_e) - \b p(\b,\l(x))\Big];\, \{\nu_x\}:
\int_{\TT^d}\!dx\int\!\nu_x(de)\,\rho(Q_e) =1 \bigg\} 
}
   \Eq(p86bis)
   $$
Since $|(1 + R_n + (\phi_k\circ |G|)_n)(1-u_n)|\le 1+k+R_n$
and, for any $\{\nu_x\}$ in the supremum above, $\int\!dx\int\!
\nu_x(de) Q_e(1+k+R_n) \le 2+k$, we can pass to the limit 
$n\to \infty$ inside the supremum and apply the Dominated 
Convergence Theorem. Then we can drop the constraint so that
\equ(p86bis) is bounded by
   $$
\eqalign{ &
\limsup_{k\to\infty}\int_{\TT^d}\!dx\,\sup_{\nu}\bigg\{
\int\!\nu(de)\,\Big[\d_0\b C_1 \big(1 + \rho(Q_e) + 
Q_e(\phi_k\circ |G|)\big)\big(1-\1_K (\rho(Q_e))\big) \cr & 
- K_{\b,\l(x)}(Q_e)\Big] - \b p(\b,\l(x)) \bigg\}
}
   \Eq(p87)
   $$
By arguing as in the proof of Lemma 4.2 of [\rcite{V}] (see also
the proof of Lemma \equ(sp4) in Section 4), since $V$ is positive 
and superstable, there is $C_2>0$ such that
   $$
|G|(\o) \le C_2 \b U(\o) 
   \Eq(p88)
   $$
where $U(\o)$ is defined as in \equ(p60) with an appropriate 
choice (depending on $|G|$) of the function $\chi$. Then, 
from \equ(p59) it follows that, for any $Q\in\MM_\t$,
   $$
Q(\phi_k\circ |G|) \le C_2 \b \Phi(Q) 
   $$
so that, setting $\d=\d_0\max\{\b C_1C_2;C_1\}$, \equ(p87) 
can be bounded by
   $$
\eqalign{&
\int_{\TT^d}\!dx\, \sup_{\nu}\int\!\nu(de)\,\Big\{ \big(
\d\b + \d\b\rho(Q_e) + \d \b \Phi(Q_e) 
- K_{\b,\l(x)}(Q_e) - \b p(\b,\l(x))\big) \cr &
\times \big(1-\1_K(\rho(Q_e))\big) 
- \big(\b p(\b,\l(x)) + K_{\b,\l(x)}(Q_e)\big) \1_K(\rho(Q_e))\Big\}
\cr &
\le \int_{\TT^d}\!dx\, \sup_m \Big\{\big(\d\b \big(1 + 
(\l(x)+\d)m\big) + \b (1-\d)p(\b (1-\d),
\l(x) +\d) \cr & 
- \b p(\b,\l(x)) - I_{\b(1-\d)}(\l(x)+\d ,m)\big)
(1-\1_K(m)) - I_\b(\l(x),m)\1_K(m) \Big\}
}
   $$
In the last inequality we used \equ(p61), \equ(p64) and introduced
the functional $I_\b(\l,m)$ defined in \equ(p39bis). Observe that, 
for any $\l\in K_u$, $m\mapsto I_\b(\l,m)$ is non negative, convex, 
and strictly positive for $m \not\in K$. Recall also that the 
pressure $p(\b,\l)$ is a continuous function of its variables. 
Then, by continuity, the superior is 0 for $\d$ (i.e. $\d_0$) 
small enough. 

\vskip.2truecm

($p=4$). Since $x\mapsto D_{\xi\eta}\l(x)$ is bounded and 
$|D_{\xi\eta}J*R_n(\o_{\cdot,\g})|\le \|DJ\|_\infty$, recalling
definitions \equ(p25) and \equ(p85bis), there is $C_4>0$ such
that 
   $$
\int_{\TT^d}\!dx\,\O_4(x) \le C_4 \int_{\TT^d}\!dx\,
\big[|G|_n(\o_{x,\g}) - k\big]^+ = C_4 \int_{\TT^d}\!dx\,
\big[|G|(\o_{x,\g}) - k\big]^+ 
   $$
Then \equ(p84bis) for $p=4$ is proved if, for $\d_0$ small enough,
   $$
\limsup_{k\to\infty}\limsup_{\g\downarrow 0}\d_0^{-1}\g^d
\log\E^{\hat f_\g}\exp\bigg[\d_0 \b C_4\g^{-d}\int_{\TT^d}\!dx\,
\big[|G|(\o_{x,\g})-k\big]^+\bigg] = 0
   \Eq(p89)
   $$

Arguing as in the proof of Theorem \equ(sp11), \equ(p89) follows if, 
for any $\d$ small enough,
   $$
\limsup_{k\to\infty}\limsup_{\g\downarrow 0} \g^d \log \E^{P_\g}
\exp\big[ T^{(\d,k)}(\o)\big] - \int_{\TT^d}\!dx\, \b p(\b,\l(x)) = 0
   \Eq(p90)
   $$
where $P_\g$ is the Poisson process on $\TT^d$ with intensity
$\g^{-d}$ and 
   $$
T^{(\d,k)}(\o) = \sum_{z\in\o} \b\l (z) - {\b\over 2} 
\sum_{\scriptstyle z,z' \in \o \atop \scriptstyle z \ne z'} 
V(\g^{-1}(z-z')) + \d \g^{-d}\int_{\TT^d}\!dx\, \big[|G|(\o_{x,\g})
-k\big]^+
   \Eq(p91)
   $$
Let $\ell_0$ be such that $U(\o) = U(\o\cap D_{\ell_0})$ with
$U$ chosen as in \equ(p88). As in the proof of Theorem \equ(sp11) 
divide $\TT^d$ into disjoint boxes $B_\s$ of side $(2\g\ell)$ and 
assume $\l(x)$ constant on this partition for small $\g$'s. Let 
$B_{\s,\ell_0}$ be as in \equ(p71bis). From \equ(p88) we can estimate:
   $$
\g^{-d}\int\limits_{\TT^d\setminus\cup B_{\s,\ell_0}}\!dx\, 
\big[|G|(\o_{x,\g})-k\big]^+ \le {\b C_2 \over 2} 
\sum_{\scriptstyle z,z' \in \o\setminus\cup B_{\s,2\ell_0} 
\atop \scriptstyle z \ne z'} V(\g^{-1}(z-z')) 
   \Eq(p92)
   $$

Restricting to $\d<C_2^{-1}$, neglecting the interaction between
different boxes and using the independence properties of $P_\g$, from
\equ(p92) we get
   $$
\E^{P_\g} \exp\big[ T^{(\d,k)}(\o)\big] \le
\prod_\s \E^{P_\g} \exp\big[T_\s^{(\d,k)}\big]
   \Eq(p93)
   $$
where 
   $$
\eqalign{
T_\s^{(\d,k)} & = \b\l(\s)N_{B_\s} - {\b\over 2} 
\sum_{\scriptstyle z,z' \in \o\cap B_\s 
\atop \scriptstyle z \ne z'} V(\g^{-1}(z-z')) \cr &
+ {\b\d C_2 \over 2} \sum_{\scriptstyle z,z' \in 
\o\cap (B_\s\setminus B_{\s,2\ell_0}) 
\atop \scriptstyle z \ne z'} V(\g^{-1}(z-z'))
+ \d \g^{-d}\int\limits_{B_{\s,\ell_0}}\!dx\, 
\big[|G|(\o_{x,\g})-k\big]^+
}
   $$
Now, expanding variables, we can rewrite:
   $$
\E^{P_\g} \exp\big[T_\s^{(\d,k)}(\o)\big] = 
Z^{(\d)}_{D_\ell}(\b,\l(\s)) \E^{\mu^{(\d,\s)}_{D_\ell}}
\exp\bigg[\d\int_{D_{\ell-\ell_0}}\!dr\, 
\big[|G|(\t_r\o)-k\big]^+\bigg]
   \Eq(p94)
   $$
where $\mu^{(\d,\s)}_{D_\ell}$ is the grand canonical
measure on $D_\ell$ with chemical potential $\l(\s)$ and
interaction energy
   $$
H_\ell(\o) - {\d C_2 \over 2} \sum_{\scriptstyle q,q' 
\in \o\cap D_{\ell,\ell_0} \atop \scriptstyle q \ne q'} V(q-q')
   $$
where $D_{\ell,\ell_0} \doteq D_\ell\setminus D_{\ell-2\ell_0}$,
the Hamiltonian $H_\ell(\o)$ was defined in \equ(p58), and
$Z^{(\d)}_{D_\ell}(\b,\l(\s))$ is the corresponding partition
function. From \equ(p93) and \equ(p94) we get
   $$
\eqalign{
\g^d\log\E^{P_\g} & \exp\big[T^{(\d,k)}(\o)\big] 
\le \g^d |D_\ell|\sum_\s |D_\ell|^{-1}
\log Z^{(\d)}_{D_\ell}(\b,\l(\s)) \cr & + 
\g^d |D_\ell| \sum_\s |D_\ell|^{-1} \log\E^{\mu^{(\d,\s)}_{D_\ell}}
\exp\bigg[\d\int_{D_{\ell-\ell_0}}\!dr\, 
\big[|G|(\t_r\o)-k\big]^+\bigg]
}
   \Eq(p95)
   $$

Since $V$ is superstable and positive, [\rcite{R2}], there is a 
positive constant $b$ such that, for any $\o\in\O$ and any 
bounded region $B\subset\R^d$, 
   $$
{1\over 2}\sum_{\scriptstyle q,q' \in \o \cap B \atop 
\scriptstyle q \ne q'} V(q-q') \ge {b N_B(\o)^2\over |B|}
   $$
Then we can estimate
   $$
\eqalign{
Z^{(\d)}_{D_\ell}(\b,\l(\s)) & \le \E^P\exp \bigg[\b \l(\s) 
N_{D_{\ell-2\ell_0}} - \b H_{\ell-2\ell_0} - {\b(1-\d C_2)b
N_{D_{\ell,\ell_0}}^2 \over |D_{\ell,\ell_0}|} + \l(\s)
N_{D_{\ell,\ell_0}} \bigg] \cr &
\le \exp \bigg[{\l(\s)^2\over 4\b b(1-\d C_2)}
|D_{\ell,\ell_0}|\bigg] Z_{D_{\ell-2\ell_0}}
(\b,\l(\s))
}
   $$
so that
   $$
\limsup_{\ell\to\infty}\limsup_{\g\downarrow 0} \g^d |D_\ell|
\sum_\s |D_\ell|^{-1} \log Z^{(\d)}_{D_\ell}(\b,\l(\s))
\le \int_{\TT^d}\!dx\, \b p(\b,\l(x))
   \Eq(p96)
   $$
On the other hand, from \equ(p88),
   $$
\E^{\mu^{(\d,\s)}_{D_\ell}}
\exp\bigg[\d\int_{D_{\ell-\ell_0}}\!dr\, 
\big[|G|(\t_r\o)-k\big]^+\bigg] \le
\E^{\mu^{(\d,\s)}_{D_\ell}}
\exp\big[\d C_2 \b H_\ell (\o)\big] 
   \Eq(p97)
   $$
and, again from superstability, the r.h.s. of \equ(p97) is
finite for any $\ell$ if $\d$ is small enough. Then, by
the Dominated Convergence Theorem,
   $$
\limsup_{k\to\infty} \E^{\mu^{(\d,\s)}_{D_\ell}}
\exp\bigg[\d\int_{D_{\ell-\ell_0}}\!dr\, 
\big[|G|(\t_r\o)-k\big]^+\bigg] =1
   $$
so that
   $$
\limsup_{\ell\to\infty}\limsup_{k\to\infty}\limsup_{\g\downarrow 0}
\g^d |D_\ell| \sum_\s |D_\ell|^{-1} \log\E^{\mu^{(\d,\s)}_{D_\ell}}
\exp\bigg[\d\int_{D_{\ell-\ell_0}}\!dr\,  
\big[|G|(\t_r\o)-k\big]^+\bigg] = 0
   \Eq(p98)
   $$
From \equ(p95), \equ(p96) and \equ(p98) we get \equ(p90) (in fact
the l.h.s. of \equ(p90) cannot be negative). \qed

\vskip.2truecm

\noindent{\it Proof of \equ(p38).} Recalling definition \equ(p33),
by Theorem \equ(sp11) the l.h.s. of \equ(p38) can be bounded by
   $$
\eqalign{
\limsup_{n\to\infty} \sup & \bigg\{ 
\int_{\TT^d}\!dx \int_{\TT^d}\!dy\, 
\Big[ \d\b \big[(\dot\l - \b|\nabla\l|^2)(s,x)
Q_x(R_nu_n) - \b \Delta\l (s,x) \cr & 
\times Q_x(P(R_n) u_n) - \b \nabla\l (s,x) \cdot \nabla J (x-y) 
Q_x(R_nu_n)Q_y(R_n) \cr &
-\b\Delta J(x-y) Q_x(P(R_n)u_n) Q_y(R_n) 
- \O(s,x,\rho(s)) Q_x(u_n) \big] \cr &
- K_{\b,\l(s,x)}(Q_x) - \b p(\b,\l(s,x)) \Big]; \{Q_x\}
\subset\MM_\t: \int_{\TT^d}\!dx\, \rho(Q_x) = 1 \bigg\}
}
   \Eq(p99)
   $$
For any $Q_x\in \MM_\t$ let $\int\!\nu_x(de)\, Q_e$ be its ergodic
decomposition. Just as argued after \equ(p86bis), we can pass to
the limit through the supremum and apply the Dominated Convergence 
Theorem. Then, calling $\bar\nu_x(dm)$ the distribution of $\rho(Q_e)$
under $\nu_x(de)$, \equ(p99) can be bounded by
   $$
\eqalign{  
\sup\bigg\{& \int\!\prod_{z\in\TT^d}\bar\nu_z(dm(z))
\int_{\TT^d}\! dx\, \Big[\d\b \big[\O(s,x,m) - \O (s,x,\rho(s)) 
\big] \1_K(m(x)) \cr & - I_\b(\l(s,x),m(x)) \Big]; \, \{\bar\nu_x\}: 
\int_{\TT^d}\!dx\int\!\bar\nu_x(dm)\, m < \infty \bigg\}
}
   \Eq(p100)
   $$
from which \equ(p38) follows immediately. \qed

\vskip.2truecm

{\bf Acknowledgments.} We thank M. Bramson, R. Esposito,
A. Kupiainen, R. Marra, and H. Spohn for discussions. 

\goodbreak
\vskip.5cm
\centerline{\bf References.} 

\vskip.2truecm
\item{[\rtag{BS}]} L. Bunimovich, Y. Sinai,
{\it Statistical properties of Lorentz gas with periodic
configuration of scatterers.}
Commun. Math. Phys. {\bf 78}, 479--497 (1981)

\vskip.2truecm
\item{[\rtag{BCS}]} L. Bunimovich, Y. Sinai, N.I. Chernov,
{\it Statistical properties of two-dimensional hyperbolic billiards.}
Russian Math. Surveys {\bf 46}, no. 4, 47--106 (1991)

\vskip.2truecm
\item{[\rtag{DP}]} A. De Masi, E. Presutti,
{\it Mathematical Methods for Hydrodynamic Limits.}
Lecture Notes in Mathematics {\bf 1501}, Springer--Verlag, Berlin 
1991 

\vskip.2truecm
\item{[\rtag{DS}]} J.D. Deuschel, D.W. Stroock,
{\it Large Deviations.}
Academic Press, Inc., Boston, MA, 1989

\vskip.2truecm
\item{[\rtag{DE}]} P. Dupuis, R.S. Ellis,
{\it A Weak Convergence Approach to the Theory of Large Deviations.}
Wiley Series in Probability and Statistics, John Wiley $\&$ Sons, 
New York 1997

\vskip.2truecm
\item{[\rtag{G1}]} H.O. Georgii,
{\it Canonical Gibbs Measures.} 
Lecture Notes in Mathematics {\bf 760}, Springer, Berlin 1979 

\vskip.2truecm
\item{[\rtag{GZ1}]} H.O. Georgii, H. Zessin, 
{\it Large deviations and the maximum entropy principle for marked 
point random fields.}
Probab. Theory Relat. Fields {\bf 96}, 177--204 (1993)

\vskip.2truecm
\item{[\rtag{GZ2}]} H.O. Georgii, H. Zessin, 
{\it Large deviations and the equivalence of ensembles for Gibbsian 
particle systems with superstable interaction.} 
Probab. Theory Relat. Fields {\bf 99}, 171--195 (1994)

\vskip.2truecm
\item{[\rtag{GL1}]} G. Giacomin, J.L. Lebowitz, {\it Phase segregation
dynamics in particle systems with long range interactions I:
macroscopic limits.} J. Stat. Phys. {\bf 87}, 37--61 (1997) 

\vskip.2truecm
\item{[\rtag{GL2}]} G. Giacomin, J.L. Lebowitz, {\it Phase segregation
dynamics in particle systems with long range interactions II: 
interface motion.} to appear on SIAM J. Appl. Math. (1998)

\vskip.2truecm
\item{[\rtag{GK}]} N. Grewe, W. Klein,
{\it The Kirkwood-Salsburg equations for bounded stable Kac potential.
II. Instability and phase transitions.}
J. Math. Phys. {\bf 18}, 1735--1740 (1977)

\vskip.2truecm
\item{[\rtag{GPV}]} M.Z. Guo, G.C. Papanicolaou, S.R.S. Varadhan,
{\it Nonlinear diffusion limit for a system with nearest neighbor 
interactions.}
Commun. Math. Phys. {\bf 118}, 31--59 (1988)

\vskip.2truecm
\item{[\rtag{KLS}]} S. Katz, J.L. Lebowitz, H. Spohn,
{\it Non equilibrium steady state of stochastic lattice gas models
of fast ionic conductors.}
J. Stat. Phys. {\bf 34}, 497--537 (1984)

\vskip.2truecm 
\item{[\rtag{K}]} W. Klein,
{\it The glass transition, possible research directions.}
Computational Materials Science {\bf 4}, 399-344 (1995)

\vskip.2truecm
\item{[\rtag{KGRCM}]} W. Klein, H. Gould, R.A. Ramos, I. Clejan,
A.I. Mel'cuk,
{\it Repulsive potentials, clumps and the metastable glass phase.}
Physica A {\bf 205}, 738--746 (1994)

\vskip.2truecm
\item{[\rtag{LPS}]} J.L. Lebowitz, E. Presutti, H. Spohn,
{\it Microscopic models of hydrodynamic behavior.}
J. Stat. Phys. {\bf 51}, 841--862 (1984)

\vskip.2truecm
\item{[\rtag{LS1}]} J.L. Lebowitz, H. Spohn,
{\it Microscopic basis for Fick's law for self-diffusion.}
J. Stat. Phys. {\bf 28}, 539--556 (1982)

\vskip.2truecm
\item{[\rtag{LS2}]} J.L. Lebowitz, H. Spohn,
{\it Steady state self-diffusion at low density.}
J. Stat. Phys. {\bf 29}, 39--55 (1982)

\vskip.2truecm
\item{[\rtag{OV}]} S. Olla, S.R.S. Varadhan,
{\it Scaling limit for interacting Ornstein-Uhlenbeck processes.}
Commun. Math. Phys. {\bf 135}, 355--378 (1991)

\vskip.2truecm
\item{[\rtag{OVY}]} S. Olla, S.R.S. Varadhan, H.T. Yau, 
{\it Hydrodynamical limit for a Hamiltonian system with weak noise.}
Commun. Math. Phys. {\bf 155}, 523--560 (1993)

\vskip.2truecm
\item{[\rtag{R}]} F. Rezakhanlou,
{\it Hydrodynamic limit for a system with finite range interactions.}
Commun. Math. Phys. {\bf 129}, 445--480 (1991)

\vskip.2truecm
\item{[\rtag{R1}]} D. Ruelle,
{\it Statistical Mechanics: Rigorous Results.}
W. A. Benjamin, Inc., New York-Amsterdam 1969 

\vskip.2truecm
\item{[\rtag{R2}]} D. Ruelle, 
{\it Superstable interactions in classical statistical mechanics.}
Commun. Math. Phys. {\bf 18}, 127--159  (1970)

\vskip.2truecm
\item{[\rtag{S}]} H. Spohn,
{\it Large Scale Dynamics of Interacting Particles.}
Text and Monographs in Physics, Springer--Verlag, Berlin 1991

\vskip.2truecm
\item{[\rtag{V}]} S.R.S. Varadhan,  
{\it Scaling limit for interacting diffusions.}
Commun. Math. Phys. {\bf 135}, 313--353 (1991)

\vskip.2truecm
\item{[\rtag{VY}]} S.R.S. Varadhan, H.T. Yau, 
{\it Diffusive limit of lattice gas with mixing conditions.}
Pre-print (1997)

\vskip.2truecm
\item{[\rtag{Y}]} H.T. Yau,  
{\it Relative entropy and hydrodynamics of Ginzburg-Landau models.}
Lett. Math. Phys. {\bf 22}, 63--80 (1991)

\end